\documentclass[11pt]{article}
\usepackage{graphicx}
\graphicspath{{./figures/}}
\usepackage{appendix}
\usepackage{latexsym,amsmath,amsfonts,amssymb,booktabs}
\usepackage[font=small]{caption}
\usepackage{slashed,upgreek,amscd,cancel,tensor,color}
\usepackage{adjustbox}
\usepackage{epsfig,latexsym,cite}
\usepackage[pdfencoding=auto]{hyperref}
\usepackage{url}
\numberwithin{equation}{section}
\usepackage{doi}

\definecolor{MyBlue}{rgb}{0.15,0.15,0.70}

\hypersetup{
colorlinks=true,
citecolor=MyBlue,
linkcolor=MyBlue,
urlcolor=MyBlue
}

\input epsf
\usepackage{amssymb}
\usepackage{amsmath}
\usepackage{amsfonts}
\usepackage{upgreek}
\usepackage{latexsym}
\usepackage{stfloats}
\usepackage{afterpage}

\newcommand\tm{\text{m}}

\newcommand{\be}{\begin{equation}}
\newcommand{\ee}{\end{equation}}
\newcommand{\beq}{\begin{equation}}
\newcommand{\eeq}{\end{equation}}
\newcommand{\bea}{\begin{eqnarray}}
\newcommand{\eea}{\end{eqnarray}}

\newcommand{\DDt}{\alpha}

\newcommand{\Q}{{\cal Q}}

\newcommand{\pid}{{\dot \pi}}

\newcommand{\gpi}{{\nabla\pi}}
\usepackage[stable]{footmisc}

\def\dkmu2{\delta K_{\mu \nu}\delta K^{\mu \nu}}
\def\pmu2{  \phi_{\mu \nu}\phi^{\mu \nu}}

\newcommand{\bs}{{\bar \sigma}}

\newcommand\deltam{{\delta}_\text{m}}
\newcommand\deltamd{\dot \delta{}_\text{m}}
\newcommand\deltamdd{\ddot \delta{}_\text{m}}

\newcommand{\al}{\alpha}

\newcommand{\om}{\omega}
\newcommand{\Om}{\Omega}

\newcommand{\tg}{\check{g}}
\newcommand{\NS}{{N_S}}

\renewcommand\[{\left[}
\renewcommand\]{\right]}

\newcommand\ees{\end{eqnarray}}
\newcommand\bees{\begin{eqnarray}}

\newcommand\alphaSm{\alpha^{\rm eff}_{\text{D,m}}}
\newcommand\alphaDm{\alpha_{\text{D,m}}}
\newcommand\alphaCm{\alpha_{\text{C,m}}}
\newcommand\alphaXm{\alpha_{\text{X,m}}}
\newcommand\alphaS{\alpha^{\rm eff}_{\text{D}}}
\newcommand\alphaH{\alpha_{\text{H}}}
\newcommand\alphaSI{\alpha^{\rm eff}_{\text{D},I}}
\newcommand\alphaDI{\alpha_{\text{D},I}}
\newcommand\alphaCI{\alpha_{\text{C},I}}

\newcommand\alphaD{\alpha_{\text{D}}}
\newcommand\alphaC{\alpha_{\text{C}}}
\newcommand\alphaX{\alpha_{\text{X}}}
\newcommand\alphaXI{\alpha_{\text{X},I}}

\newcommand\alphaB{\alpha_{\text{B}}}
\newcommand\alphaM{\alpha_{\text{M}}}
\newcommand\alphaK{\alpha_{\text{K}}}
\newcommand\alphaT{\alpha_{\text{T}}}
\newcommand\cI{c_{s,I}}

\newcommand{\Psiosc}{\Psi_\mathrm{osc}}
\newcommand{\Phiosc}{\Phi_\mathrm{osc}}

\newcommand{\vosc}{v_\mathrm{osc}}

\newcommand{\OO}{\mathcal{O}}

\begin{document}
\hfill CERN-TH-2016-192
\vspace{0.5cm}

\begin{center}
\Large{\textbf{Weakening Gravity on Redshift-Survey Scales \\[0.2cm] with Kinetic Matter Mixing  }} \\[1cm]

\large{Guido D'Amico$^{\rm a}$, Zhiqi Huang$^{\rm b}$,\\[0.1cm]
Michele Mancarella$^{\rm c, d}$ and Filippo Vernizzi$^{\rm c, d}$}
\\[0.5cm]

\small{
\textit{$^{\rm a}$  Theoretical Physics Department, CERN, Geneva, Switzerland}}
\vspace{.2cm}

\small{
\textit{$^{\rm b}$  School of Physics and Astronomy, Sun Yat-Sen University, \\ [0.05cm]
135 Xingang Xi Road, 510275, Guangzhou, China}}
\vspace{.2cm}

\small{
\textit{$^{\rm c}$ CEA, IPhT, 91191 Gif-sur-Yvette c\'edex, France \\ [0.05cm]
CNRS,  URA-2306, 91191 Gif-sur-Yvette c\'edex, France}}

\vspace{.2cm}

\small{
\textit{$^{\rm d}$ Universit\'e Paris Sud, 15 rue George Cl\'emenceau, 91405,  Orsay, France}}

\vspace{.2cm}

\vspace{0.5cm}
\today

\end{center}

\vspace{2cm}

\begin{abstract}

We explore general scalar-tensor models in the presence of a kinetic mixing between matter and the scalar field, which we call Kinetic Matter Mixing.
In the frame where gravity is de-mixed from the scalar this is due to disformal couplings of matter species to the gravitational sector, with disformal coefficients that depend on the gradient of the scalar field. In the frame where matter is minimally coupled, it originates from the so-called beyond Horndeski quadratic Lagrangian. We extend the Effective Theory of Interacting Dark Energy by allowing disformal coupling coefficients to depend on the gradient of the scalar field as well. In this very general approach, we derive the conditions to avoid ghost and gradient instabilities and we define Kinetic Matter Mixing independently of the frame metric used to described the action.
We study its phenomenological consequences  for a $\Lambda$CDM background evolution, first analytically on small scales. Then, we compute the matter power spectrum and the angular spectra of the CMB anisotropies and the CMB lensing potential, on all scales. We employ the public version of COOP, a numerical Einstein-Boltzmann solver that implements very general scalar-tensor modifications of gravity.
Rather uniquely, Kinetic Matter Mixing weakens gravity on short scales, predicting a lower $\sigma_8$ with respect to the $\Lambda$CDM case.
We propose this as a possible solution to the tension between the CMB best-fit model and low-redshift observables.

\end{abstract}

\newpage

\tableofcontents

\vspace{.5cm}
\section{Introduction}

A key goal of current and future cosmic surveys is to
constrain or possibly detect deviations from the standard  $\Lambda$CDM model, which are expected if the origin of  the present  accelerated expansion is not a cosmological constant, but a dynamical field or a modification of General Relativity (see e.g. \cite{Amendola:2012ys,Amendola:2016saw}).
To deal with the fact that there are many dark energy and modified gravity models (see for instance \cite{Clifton:2011jh,Joyce:2014kja}), effective approaches that describe these  deviations  for a large number of models in terms of a few time-dependent parameters have been proposed in the literature \cite{Creminelli:2008wc,Gubitosi:2012hu,Bloomfield:2012ff,Gleyzes:2013ooa,Bloomfield:2013efa,Gleyzes:2014rba,Battye:2012eu,Battye:2013ida,Baker:2011jy,Baker:2012zs,Skordis:2015yra,Lagos:2016wyv}.
In most cases, these approaches are limited to a description of  cosmological perturbations around a Friedmann-Lema\^itre-Robertson-Walker (FLRW) background in the linear regime (see however \cite{Bellini:2015wfa,Bellini:2015oua} for some nonlinear aspects), applicable to scales above $\sim 10$Mpc, where deviations from General Relativity are not yet well tested.

This work focuses on the so-called Effective Theory of  Dark Energy.
Formulated for  single scalar field models---i.e.~models where the time diffeomorphisms are broken while leaving the spatial ones preserved---in this approach the unitary (or uniform field) gauge action is given as the sum of all possible geometrical elements constructed from the metric and its derivatives that are invariant under the preserved diffs, i.e.~the spatial ones \cite{Creminelli:2006xe,Cheung:2007st}. It has been derived and studied for minimally and nonminimally coupled dark energy models, respectively, in \cite{Creminelli:2008wc} and \cite{Gubitosi:2012hu,Gleyzes:2013ooa} (see \cite{Piazza:2013coa,Tsujikawa:2014mba,Gleyzes:2014rba} for reviews). When restricting to the lowest order in derivatives, the final second-order action contains five free functions of time that parametrize any deviation from $\Lambda$CDM. As shown in \cite{Gleyzes:2013ooa}, four of these functions describe cosmological perturbations
of effective theories of dark energy or modified gravity within the Horndeski class, i.e.~those with quadratic gravitational  action with the same structure as Horndeski theories \cite{Horndeski:1974wa,Deffayet:2011gz,Kobayashi:2011nu}.
This description has been reformulated in \cite{Bellini:2014fua} in terms of dimensionless functions that clearly parametrize deviations from General Relativity.
The fifth function, denoted as $\alphaH$, describes scalar field models extending the Horndeski class, such as, e.g.,~the theories ``beyond Horndeski'' proposed in \cite{Gleyzes:2014dya,Gleyzes:2014qga} (see \cite{Zumalacarregui:2013pma} for an earlier proposal of theories beyond Horndeski).
The Effective Theory formulation has been used to explore the observational consequences of deviations from $\Lambda$CDM (see for instance \cite{Piazza:2013pua,Kase:2014yya,Ade:2015rim,Lombriser:2015cla,Perenon:2015sla,Gleyzes:2015rua,Frusciante:2016xoj,Hu:2016zrh,Salvatelli:2016mgy,Renk:2016olm,Leung:2016xli,Pogosian:2016pwr}).
In this direction, a few Einstein-Boltzmann solvers have been recently developed and employed \cite{Hu:2013twa,Raveri:2014cka,Bellini:2015xja,Zumalacarregui:2016pph,Huang:2015srv,zqhuang_2016_61166}.

References~\cite{Gubitosi:2012hu,Gleyzes:2013ooa} assumed that   all matter species are minimally coupled to the same metric, which  we call Jordan frame metric for convenience.
In general, however, there is no reason to impose this restriction. The universality of couplings is very well tested on Solar System scales \cite{Will:2014xja} but on cosmological scales constraints are much weaker and different species could have distinct couplings to the gravitational sector.
If matter is universally but {\em nonminimally} coupled to the gravitational sector, in most cases
it is convenient to  perform a field redefinition of the metric that brings the system into the Jordan frame, where matter is minimally coupled.
In general, this frame transformation depends on the scalar field and its derivatives and, as long as it is regular and invertible, it cannot change the physics (see e.g.~\cite{Domenech:2015tca}).  The  advantage of using  the Jordan frame to derive predictions is that only the gravitational sector is non-standard; thus, one does not need to care about modifications of non-gravitational forces, which would otherwise greatly complicate the analysis.

Along this line of thought, recently Ref.~\cite{Gleyzes:2015pma}  extended the effective approach of \cite{Gubitosi:2012hu,Gleyzes:2013ooa} to allow for distinct conformal and disformal couplings of matter species to the gravitational sector.
The treatment was restricted to effective theories within the Horndeski class and to conformal and disformal factors that depend only on the scalar field (not on its gradients).
In this case, the full quadratic action depends on the four functions describing the gravitational sector and on two extra functions per species, describing the coupling to the scalar. However, two of these functions are redundant, because the structure of the action is preserved under  transformations of the reference metric. This is expected, as it was shown  that the structure of the Horndeski Lagrangians is  preserved under disformal transformations  with both conformal and disformal coefficients independent of the scalar field gradient \cite{Bettoni:2013diz}.

The phenomenological aspects of general modifications of gravity described by Ref.~\cite{Gleyzes:2015pma} was studied
in Ref.~\cite{Gleyzes:2015rua}, where  constraints on the effective descriptions were derived from three observables: the galaxy and  weak-lensing power spectra and the correlation between the Integrated Sachs-Wolfe (ISW) effect and the galaxy distribution. However, the study was restricted to the quasi-static limit, which is  reliable on short enough scales and at late times, once the oscillations of the scalar fluctuations have been damped by the expansion of the universe.
While this approximation is fairly good  for current and future galaxy and weak lensing surveys,\footnote{The quasi-static approximation typically fails on scales $k \lesssim  a H /c_s$, where $c_s$ is the sound speed of fluctuations of the scalar. As shown in \cite{Sawicki:2015zya}, this approximation should be reliable  for surveys such as Euclid  as long as the sound speed exceeds $10 \%$ of the speed of light, i.e.~$c_s \gtrsim 0.1$.} it fails on large scales or high redshifts.

In this article we go one step forward, in two directions. First, in Sec.~\ref{section2} we extend the treatment of  Ref.~\cite{Gleyzes:2015pma} and include in the gravitational action  the fifth time-dependent function, $\alphaH$, describing models extending the Horndeski class.
As shown in  \cite{Gleyzes:2014dya,Gleyzes:2014qga}, the structure of the Lagrangian of theories beyond Horndeski is preserved under a disformal transformation of the metric with disformal coefficient that depends
as well on the {\em gradient} of the scalar field, i.e.~of the form
\be
\label{disftr}
\tilde g_{\mu \nu} = C(\phi) g_{\mu \nu} + D(\phi,X) \partial_\mu \phi \partial_\nu \phi  \;, \qquad X \equiv g^{\mu \nu} \partial_\mu \phi \partial_\nu \phi  \;.
\ee
Thus, in the following we consider the possibility that matter  couples to a  Jordan frame metric of this form.\footnote{Disformal transformations with $C=C(\phi,X)$ have been studied in the context of beyond Horndeski theories in \cite{Zumalacarregui:2013pma,Bettoni:2015wta} and in the context of degenerate higher-order theories in \cite{Crisostomi:2016tcp,Langlois:2015cwa,Achour:2016rkg,Crisostomi:2016czh}.} In particular, we  denote  the conformal and disformal coefficients of the nonminimal coupling of matter respectively as $C_{\rm m}(\phi)$ and $D_{\rm m} (\phi,X)$ (which can be distinct for different species).

As shown in Sec.~\ref{section2},
the dependence of the disformal coupling on the derivative of the field introduces a kinetic mixing between the scalar and matter, which hereafter  we call \emph{Kinetic Matter Mixing} (KMM), that has rather unique  observational effects, as discussed below.
To parametrize this direct kinetic coupling we introduce an additional function of time,
\be
\alpha_{\rm X, m} = \frac{X^2}{C_{\rm m}} \frac{\partial D_{\rm m}}{\partial X} \;,
\ee
where the right-hand side is evaluated on the background.
Thus, the full quadratic action depends now on five functions describing the gravitational sector and three functions per species, describing the matter couplings. The structure of this action is preserved under transformations of the reference metric of the form \eqref{disftr}. Remarkably,
$\alpha_{\rm X, m} $ is transformed into the beyond Horndeski parameter $\alphaH$ under a  transformation which sets to zero the disformal coupling.
Since KMM  is a truely physical effect, it is possible to define  a combination of these two parameters, proportional to $\left(\alphaH - \alpha_{\rm X, m}  \right)^2 $ (c.f.~eq.~\eqref{gpar} below), that encodes in a frame-independent way the degree of kinetic mixing between matter and the scalar.

While in Sec.~\ref{section2} we assume for simplicity that matter couples universally to the same Jordan frame metric, in App.~\ref{Lagmatter} we extend this treatment to multiple species with distinct couplings. Taking into account the invariance under the  disformal transformation \eqref{disftr}, which reduces the number of independent functions of time by three, the whole system
depends  on a total of $2  + 3 N_S $ independent functions of time, where  $N_S $ is the number of  matter species.

In the rest of the paper we assume that matter is universally coupled to the gravitational sector and work in the Jordan frame, where the coupling is minimal.
In this frame, KMM is encoded in the beyond Horndeski parameter $\alphaH$.
We then extend the treatment of Ref.~\cite{Gleyzes:2015rua} and explore the phenomenological consequences of general late-time modifications of gravity  including beyond Horndeski theories (see also \cite{Kobayashi:2014ida,DeFelice:2015isa} and \cite{Saito:2015fza,Kase:2015zva,Sakstein:2016ggl,Babichev:2016jom} for an earlier study of the observational consequences of beyond Horndeski theories, respectively in cosmology and astrophysics).
In Sec.~\ref{sec3}, we focus on short scales.
In particular, we derive the eigenmodes of propagation of the scalar field and matter, which in the presence of a nonvanishing $\alphaH$ are mixed by their kinetic coupling. Moreover, we obtain the evolution equations in the quasi-static regime, which govern the dynamics once the oscillating modes have been damped by the expansion. Appendix \ref{Fullaction} contains the full action of perturbations in Newtonian gauge, derived for completeness, while the transition between the oscillating regime and the quasi-static limit is discussed in App.~\ref{app:QS}.

In Sec.~\ref{sec:alphaH} we go beyond the quasi-static approximation and explore the full range of cosmological scales using the linear Einstein-Boltzmann solver of Cosmology Object Oriented Package (COOP)~\cite{zqhuang_2016_61166},\footnote{See \url{http://www.cita.utoronto.ca/~zqhuang/} for documentation.} which solves cosmological perturbations
including very general deviations from $\Lambda$CDM in terms of the Effective Theory of Dark Energy description \cite{Gleyzes:2014rba}. In particular,  assuming the background expansion history of $\Lambda$CDM, we compute the matter power spectrum, the Cosmic Microwave Background (CMB) anisotropies angular power spectrum, and the CMB lensing potential angular spectrum in the presence of KMM, for a non vanishing $\alphaH$ parameter. As we will see, on ``short'' scales, i.e.~for $k \gtrsim 10^{-3}h\,\text{Mpc}^{-1}$, the quasi-static approximation provides the  correct amplitude for the linear growth factor, which is scale independent and suppressed with respect to the $\Lambda$CDM case.  On larger scales, we compute the linear matter growth analytically using a perturbative expansion in $\alphaH$ that confirms the numerical results.
To contrast with the effects of $\alphaH$, in App.~\ref{sec:alphaB} we compute the same observables in the case of a kinetic mixing between the scalar field and gravity, the so called kinetic braiding~\cite{Deffayet:2010qz,Pujolas:2011he} (see \cite{Creminelli:2006xe,Creminelli:2008wc} for an earlier study), and we find agreement with the results of Ref.~\cite{Zumalacarregui:2016pph}. We compare these results with the quasi-static approximation and a perturbative expansion in the braiding parameter. In contrast to kinetic braiding or other modifications of gravity within the Horndeski class,
the exchange of fifth force in KMM suppresses the power of matter perturbations on redshift-survey scales.
In Sec.~\ref{sec4.3}, we study the possibility that the lack of power measured in the large scale structures and in tension with that inferred from the CMB anisotropies observed by Planck~\cite{Ade:2013zuv,Ade:2015xua} can be explained by the KMM special signature. Finally, we conclude in Sec.~\ref{sec_last}.


\section{Effective Theory of Dark Energy with Kinetic Matter Mixing}
\label{section2}

In this section we extend the treatment of \cite{Gleyzes:2015pma}, limited to Horndeski theories, and develop the unifying framework for dark energy and modified gravity that allows distinct conformal-disformal couplings of matter species to the gravitational sector, including beyond Horndeski theories. We show that the quadratic beyond Horndeski operator  arises when transforming to the Jordan frame a disformal coupling of matter species which depends on the kinetic energy of the scalar field. In this setup, we derive the conditions to avoid ghost and gradient instabilities and discuss the disformal/conformal transformations of the gravitational and matter action. The reader only interested  in the phenomenological aspects of KMM is invited to skip this section and go directly to Sec.~\ref{sec3}, not before having retained eq.~\eqref{S2} as the second order action describing the gravitational sector.

\subsection{Gravitational and matter actions}
\label{sec2}

In the present work, following \cite{Gubitosi:2012hu,Gleyzes:2013ooa} we assume that the gravitational sector is described by a four-dimensional metric $g_{\mu \nu}$ and a scalar field $\phi$.
As usual, we choose a coordinate system such that the constant time hypersurfaces coincide with the  uniform scalar field hypersurfaces.
In this gauge, referred to  as unitary gauge, the metric  can be written in the ADM form,
\be
\label{ADM}
ds^2=-N^2 dt^2 +{h}_{ij}\left(dx^i + N^i dt\right)\left(dx^j + N^j dt\right) \, ,
\ee
where $N$ is the lapse and $N^i$ the shift.  In the following,  a dot stands for a time derivative with respect to $t$, and $D_i$ denotes the covariant derivative associated with the three-dimensional spatial metric $h_{ij}$. Spatial indices are lowered and raised with the spatial metric $h_{ij}$ or its inverse $h^{ij}$, respectively.

In the unitary gauge, a generic gravitational action  can be written in terms of  geometric quantities that are invariant under spatial diffeomorphisms \cite{Creminelli:2006xe,Cheung:2007st}. Expressed in the ADM coordinates introduced above, these geometric quantities are the lapse $N$,
the extrinsic curvature  of the constant time hypersurfaces $K_{ij}$,
whose components are given by
 \be
\label{extrinsic_ADM}
K_{ij} = \frac{1}{2N} \big(\dot h_{ij} - D_i N_j - D_j N_i \big) \;,
\ee
as well as  the
3d Ricci tensor of the constant time hypersurfaces $R_{ij}$
and, possibly,  spatial derivatives of all these quantities. Thus, the gravitational action is generically of the form
 \beq
\label{g_action}
S_{\rm g}=\int d^4 x \sqrt{-g}\,  L(N,  K_{ij}, R_{ij}, h_{ij},D_i ;t) \;.
\eeq

To study linear perturbations, one needs to expand the action at second order around a homogeneous background. For the background geometry, we assume  a spatially flat FLRW metric, $ds^2 = - dt^2 + a^2(t) d \vec x^2$.  Its dynamics is governed by the background evolution equations and we refer the reader to  Refs.~\cite{Gleyzes:2013ooa,Gleyzes:2014rba,Gleyzes:2015pma} for details on their derivation.
We can now expand the gravitational action up to second order in perturbations.
Fixing the background gauge $\bar N=1$,  these are
\be
\delta N = N - 1 \,, \qquad \delta K_{ij} = K_{ij} - H h_{ij}\, ,
\ee
as well as $R_{ij}$, which is already a perturbation since its background value vanishes. It is convenient to introduce the time-dependent parameters $\alphaK$, $\alphaB$, $\alphaT$ \cite{Bellini:2014fua} and $\alpha_{\rm H}$ \cite{Gleyzes:2014qga}
in terms of which the second-order gravitational action reads
\be
\begin{split}
\label{S2}
 S_{\rm g}^{(2)}=  \int d^3x dt \,a^3   \frac{M^2}{2}   \bigg[ & \delta K^i_j \delta K^j_i-\delta K^2 + (1+\alphaH) R {\delta N}{}+(1+\alphaT) \, \delta_2 \Big(   {\sqrt{h}}R/{a^3 }\Big)    \\
 &   +  \alphaK H^2 {\delta N}^2 + 4 \alphaB H \delta K {\delta N}{}  \bigg]   \, ,
 \end{split}
 \ee
where $\delta_2$ denotes taking the expansion at second order in perturbations.
Another useful parameter is the variation of the  effective Planck mass squared  $M^2$,
\beq
\label{alphaM}
\alphaM\equiv \frac{d \ln M^2}{ d \ln a}\,.
\eeq
For the details on the derivation of the above action and the explicit definitions of the  parameters $\alphaK$, $\alphaB$, $\alphaM$, $\alphaT$ and $\alphaH$ in terms of  first and second derivatives of $L$ with respect to its arguments, we refer again the reader to Refs.~\cite{Gleyzes:2013ooa,Gleyzes:2014rba,Gleyzes:2015pma}.

The gravitational action must be supplemented by a matter action $S_{\rm m}$,
\be
S_{\rm m} =   \int d^4 x \sqrt{- \tg }\,  L_I \Big(   \tg_{\mu \nu}, \psi \Big) \; ,
\ee
where $\tg_{\mu \nu}$ is the Jordan-frame metric.
In order to describe dark energy and modified gravity scenarios where the scalar and matter can be kinetically mixed, we assume that this metric is conformally and disformally related to the gravitational metric $g_{\mu \nu}$ by
\be
\label{disf_unit_I2_m}
\tg_{\mu \nu} = C_{\rm m}^{(\phi)}(\phi) g_{\mu \nu}  + D^{(\phi)}_{\rm m}(\phi, X) \partial_\mu \phi \, \partial_\nu \phi \;.
\ee
Contrarily to the disformal coupling presented in \cite{Gleyzes:2015pma}, here $D^{(\phi)}_{\rm m}$ can also depend on $X$, to allow for a kinetic mixing.
In Appendix~\ref{Lagmatter} we generalize to the case where matter is made of several species, each of which is coupled to a different metric.

To conclude, we notice that the  variation of the matter action $S_{\rm m}$  with respect to the metric $g_{\mu\nu}$ defines the energy-momentum tensor, according to  the standard expression
 \be
T^{\mu\nu} \equiv  \frac{2}{\sqrt{-g}} \frac{\delta S_{\rm m}}{\delta g_{\mu\nu}}\;.
\ee
This definition applies even if matter is minimally coupled to a metric $\tg_{\mu\nu}$ that differs from $g_{\mu\nu}$. In the homogeneous case, the energy-momentum tensor depends only on the energy density $\rho_{\rm m} \equiv -\bar T^{0}_{\ 0}$ and  the pressure $p_{\rm m} \equiv \bar T^i_{\ i}/3$.

\subsection{Matter couplings and stability conditions}
\label{sec:ma}

To discuss the stability and determine the propagation speed of dark energy perturbations,  one must also include quadratic terms  that come from the matter action, because the latter depends on the gravitational degrees of freedom. In order to do so, we need to  take into account  that   matter is minimally coupled to a metric $\tg_{\mu \nu} $ defined in eq.~\eqref{disf_unit_I2_m}.

In unitary gauge,  this definition  reads
\be
\label{disf_unit_I_m}
\tg_{\mu \nu} = C_{\rm m}(t) g_{\mu \nu}  + D_{\rm m}(t , N) \delta_\mu^0 \delta_\nu^0 \;,
\ee
with
\be
C_{\rm m}(t) =  C_{\rm m}^{(\phi)} \big( \phi( t) \big)\, , \qquad D_{\rm m}(t, N) =   \dot{ \phi}^2 (t) D_{\rm m}^{(\phi)} \big( \phi( t) , -\dot \phi( t)^2/N^2 \big)\,.
\ee
Then, we introduce the parameters
\be
\label{defalphas_m}
 \alphaCm \equiv \frac{\dot C_{\rm m} }{2 H  C_{\rm m}} \, , \qquad \alphaDm \equiv \frac{D_{\rm m}}{  C_{\rm m}-D_{\rm m}}\, , \qquad \alphaXm \equiv - \frac{1}{2  C_{\rm m} } \frac{\partial D_{\rm m} }{\partial N} \;  ,
 \ee
 where the right-hand sides are evaluated on the background.
 The first two parameters in the above equations, $\alphaCm$ and $\alphaDm$, were introduced in Ref.~\cite{Gleyzes:2015pma}.

Combining the quadratic action for matter with eq.~\eqref{S2}, one can extract  the dynamics of the gravitational scalar degree of freedom and the matter ones. The explicit calculation in the case of perfect fluids is presented in Appendix~\ref{Lagmatter}. The absence of ghosts is guaranteed by the positivity of the matrix in front of the kinetic terms. For the gravitational scalar degree of freedom, this condition is given by

\be
\DDt \equiv \alphaK + 6 \alphaB^2 +3 \alphaSm \, \Omega_{\rm m} \geq0\; , \label{alpha_def}
\ee
where
$\Omega_{\rm m}$
is the standard (time-dependent) dimensionless density  parameter,
\be
\Omega_{\rm m}\equiv \frac{\rho_{\rm m}}{3M^2 H^2}\,,
\label{Omegam}
\ee
and we define the combination
\be
\label{alphaSI}
\alphaSm \equiv \alphaDm(1+\alphaXm)^2+\alphaXm(2+\alphaXm)+ \frac{1}{2 C_{\rm m}} \frac{\partial^2  D_{\rm m}}{\partial N^2} \;.
\ee
Thus, the dependence on $X$ in the disformal coupling affects the ghost-free condition.

Diagonalization of the kinetic matrix yields the following dispersion relation (see Appendix~\ref{Lagmatter} for a generalization to multiple species)
\begin{equation}
( \omega^2 -  c_s^2 k^2) (\omega^2 - c_{\rm m}^2 k^2)  =  \lambda^2 c_s^2    \, \omega^2 k^2    \, , \label{km}
\end{equation}
where the parameter $\lambda^2$ on the right-hand side is defined as
\be
\lambda^2 \equiv \frac{3}{\alpha c_s^2}     \Big[1+(1+\alphaDm)w_{\rm m} \Big] \Omega_{\rm m} \, (\alphaH-\alphaXm)^2  \; .
\label{gpar}
\ee
This is the physically relevant parameter measuring the degree of KMM (as expected it is frame independent, see below).
The $c_s^2$ appearing above is the sound speed of dark energy for $\lambda=0$, given by
\be
\label{tildecss}
\begin{split}
c_s^2 =  - \frac{1}{\alpha} \bigg\{ & 2(1+\alphaB) \bigg[ \xi  +(1+\alphaH) \frac{\dot H}{H^2} -  \frac{ \dot \alpha_{\rm H}}{H} \bigg] + 2 \frac{ \dot \alpha_{\rm B} }{H}  \\
&+  3 (1+\alphaH)^2  \left[ 1 +(1+\alphaDm)  w_{\rm m}\right] \Omega_{\rm m}  \bigg\} \;,
\end{split}
\ee
where for convenience we have defined
\be
\xi \equiv \alphaB (1+\alphaT) + \alphaT - \alphaM - \alphaH (1+\alphaM) \;. \label{xidef}
\ee
For $\alphaH=0$,  this coincides with the parameter $\xi$ first defined in \cite{Gleyzes:2015pma}.
The above dispersion relation yields the two speeds of propagation
\be
\label{cpm}
c_\pm^2 = \frac12 \Big\{ c_{\rm m}^2 + c_s^2 (1+\lambda^2) \pm \sqrt{\big[ c_{\rm m}^2 + c_s^2 (1+\lambda^2)\big]^2 - 4 c_{\rm m}^2 c_s^2}  \Big\} \; .
\ee
Equations \eqref{km} and \eqref{cpm} generalize the dispersion relations and speeds of propagation derived in \cite{Gergely:2014rna,Gleyzes:2014dya,DeFelice:2015isa} for $\alphaXm=0$.
The effect of KMM appears in the presence of the coupling $\lambda^2 \neq 0$
and the propagation modes are mixed states of matter and scalar.
In general, absence of gradient instabilities is guaranteed by the usual conditions $ c_\pm^2 \geq 0$.
Finally, when $\alphaXm = \alphaH $ we recover the usual results, i.e.~$c_+^2 = c_s^2$ and $c_-^2 = c_{\rm m}^2$ for $c_s^2>c_{\rm m}^2$.

\subsection{\texorpdfstring{$(\partial \phi)^2$}{(\partial \phi)^2}-dependent disformal transformations}
\label{sec:disf}
As mentioned earlier, there is  some arbitrariness in the choice of the metric $g_{\mu\nu}$ that describes the gravitational sector. Let us thus see how the description is modified when the reference metric undergoes a disformal transformation, of the form
\be
g_{\mu \nu} \to \tilde g_{\mu \nu} = C^{(\phi)}( \phi ) g_{\mu \nu} +  D^{(\phi)} ( \phi , X) \partial_\mu \phi \partial_\nu \phi \;, \label{disf_trans}
\ee
which in unitary gauge corresponds to
\be
\label{disf_uni}
g_{\mu \nu} \to \tilde g_{\mu \nu} = C(t) g_{\mu \nu}  + D(t, N) \delta_\mu^0 \delta_\nu^0  \;.
\ee
The effect of this transformation on the ADM variables, on the background quantities and on the linear perturbations  has been studied in detail in \cite{Gleyzes:2014qga,Gleyzes:2015pma}. Here, we  present the main consequences on the parametrization of the gravitational sector.

In analogy with \eqref{defalphas_m}, it is convenient to introduce the dimensionless time-dependent parameters
\be
\label{defalphas2}
 \alphaC \equiv \frac{\dot C}{2 H  C} \, , \qquad \alphaD \equiv \frac{D}{ C-D}\, , \qquad \alphaX \equiv - \frac{1}{2 C } \frac{\partial D}{\partial N}\; ,
\ee
 which characterize the conformal and disformal parts of the above metric transformation.\footnote{As for the transformations in Secs.~\ref{sec2} and \ref{sec:ma}, we require $C>0$ and $\alphaD > -1$.}

Let us first see how  the gravitational  action \eqref{S2} changes under the transformation (\ref{disf_uni}).
As shown in Ref.~\cite{Gleyzes:2014qga}, the structure of the combination of the Horndeski and  beyond Horndeski Lagrangians is preserved under a disformal transformation with an $X$-dependent disformal function $D$. Indeed, one can check that  \eqref{S2} maintains the same structure with
the time-dependent coefficients in the action transforming as
\be
 \tilde{M}^2 =\frac{M^2 }{ C\sqrt{1+\alphaD}}\, \label{Mtilde}
\ee
and
\be
\begin{split}
\label{alphatilde}
\tilde{\alpha}_{\rm K} &=\frac{\alphaK+12\alphaB \alpha_{\rm CDX}-6  \alpha_{\rm CDX}^2 +3 \Omega_{\rm m} (1+\alphaXm) \alphaS}{(1+\alpha_{\rm CDX})^2} \;, \\
\tilde{\alpha}_{\rm B}&= \frac{1+\alphaB}{1+\alpha_{\rm CDX}}-1 \;,  \\
\tilde{\alpha}_{\rm M} &= \frac{\alphaM- 2 \alphaC }{1+\alphaC} - \frac{\dot \alpha _{\rm D}}{2 H  (1+\alphaD) (1+\alphaC)} \;, \\
\tilde{\alpha}_{\rm T}&=(1+\alphaT)(1+\alphaD)-1\; , \\
\tilde{\alpha}_{\rm H}&= \frac{\alphaH - \alphaX}{1+\alphaX} \;,
\end{split}
\ee
where $\alpha_{\rm CDX} \equiv  (1+\alphaC)(1+\alphaD)(1+\alphaX) -1 $
and, in analogy with the definition \eqref{alphaSI}, we have introduced
\be
\alphaS \equiv \alphaD(1+\alphaX)^2+\alphaX(2+\alphaX)+ \frac{1}{2 C} \frac{\partial^2  D}{\partial N^2} \;.
\ee
We can use these transformations, which depend on the three arbitrary functions $\alphaC$, $\alphaD$ and $\alphaX$, to set to zero any three of the parameters $\tilde\alpha_a$ above.

Finally, the conformal and disformal coefficients associated with the respective matter Jordan frame metrics are modified according to
\be
\label{alphaDC_change}
\begin{split}
\tilde{\alpha}_{\text{D,m}} & = \frac{\alphaDm - \alphaD}{1+\alphaD} \;,  \\
\tilde{\alpha}_{\text{C,m}}  &= \frac{\alphaCm - \alphaC}{1+\alphaC} \; ,\\
\tilde{\alpha}_{\text{X,m}}  &= \frac{\alphaXm - \alphaX}{1+\alphaX} \; .
\end{split}
\ee
These transformations can be straightforwardly extended to the case of different couplings to different species, for instance by simply replacing $\tilde{\alpha}_{\text{D,m}}$ by $\tilde{\alpha}_{\text{D},I}$ and $\alphaDm$ by $\alphaDI$.

One can verify that  the stability condition \eqref{alpha_def} is frame independent. In particular, $\alpha$ transforms as
\be
\tilde \alpha= \frac{\alpha}{(1+\alpha_{\rm CDX})^2}  \, .
\ee
It is also  straightforward to check that all the propagation speeds, i.e.~of tensor, scalar and matter fluctuations, transform in the same way and that their signs remain unchanged,
\be
\label{changecs}
 \tilde c_T^2 = (1+\alphaD)c_T^2\,,\quad \tilde c_s^2 = (1+\alphaD)c_s^2\,, \quad \tilde c_{\rm m}^2 =  (1+\alphaD)c_{\rm m}^2\,.
 \ee
Finally, using these expressions and those in \cite{Gleyzes:2015pma} it is possible to show that the parameter $\lambda^2$ defined in eq.~\eqref{gpar}, which measures the degree of KMM, is frame independent as expected.

\section{Short-scale dynamics}
\label{sec3}

In this section we discuss the short-scale dynamics of cosmological perturbations. We assume universal coupling of matter species and, without loss of generality, minimal coupling. Thus, the action describing perturbations is given by \eqref{S2}, where the gravitational metric $g_{\mu \nu}$ is the Jordan frame metric.
We focus on the scalar fluctuations and we employ the usual Stueckelberg procedure\cite{Cheung:2007st}, $t \to t + \pi(t ,\vec x)$, to move from the unitary gauge to the Newtonian gauge, whose metric for a flat FLRW universe reads
\be
ds^2= - (1+2 \Phi) dt^2+a^2 (1-2 \Psi) d\vec{x}^2\; .
\label{Newtoniangauge}
\ee

On short scales, the gradients of the scalar field $\phi$ support an oscillatory regime. In the presence of KMM, i.e.~$\lambda^2 \neq 0$, the oscillations are also shared by matter, even when matter is made of nonrelativistic species with no pressure gradients. We first describe these oscillations and their normal modes in the next subsection, while in Sec.~\ref{sec:QS} we discuss the late-time quasi-static regime occurring after the oscillations decay.

\subsection{Oscillatory regime and normal modes}

In this subsection, to describe matter we use a derivatively coupled scalar field $\sigma$, with action
\be
\label{Smatter}
S_{\rm m}=\! \int \!d^4x \sqrt{-g}\, P(Y), \, \quad Y\equiv g^{\mu\nu}\partial_{\mu}\sigma\partial_{\nu}\sigma \; ,
\ee
and we define the background energy density and pressure and the matter sound speed respectively as
\be
\rho_{\rm m}=-2 \dot{\sigma}_0^2 P_Y(Y)-P(Y),  \qquad p_{\rm m}=P(Y)\; , \qquad c_{\rm m}^2\equiv\frac{P_Y}{P_Y-2 \dot{\sigma}_0^2 P_{YY}}\; .
\ee
We also introduce the energy density contrast and the velocity potential respectively as
\be
\deltam \equiv \frac{\delta \rho_{\rm m}}{\rho_{\rm m}} \;, \qquad v_{\rm m} \equiv - \frac{\delta \sigma}{\dot \sigma_0}\;.
\label{defdeltav}
\ee
For completeness, the full second-order actions describing the gravitational and matter sectors in Newtonian gauge in this case are given in Appendix \ref{Fullaction}, eqs.~\eqref{fullacgrav} and \eqref{fullacmat}.

To study the normal modes of oscillations we consider the kinetic limit, i.e.~the limit where the spatial and time derivatives are larger than the expansion rate $H$. In this case,  it is possible to find a  redefinition of the metric perturbations  that  de-mixes  the new metric variables  from the scalar field $\pi$ and removes the higher derivative term from the gravitational action.  This is explicitly given by \cite{Gleyzes:2014qga}
\be
\label{toE}
\begin{split}
\Phi_E & \equiv     \frac{1+\alphaH}{1+\alphaT}\Phi +   \bigg( \frac{1+\alphaM}{1+\alphaT} - \frac{1+  \alphaB}{1+\alphaH} \bigg) H \pi - \frac{\alphaH}{{1+\alphaT}} \pid  \; , \\
\Psi_E & \equiv \Psi+  \frac{\alphaH-\alphaB}{1+\alphaH} H \pi \, .
\end{split}
\ee
Using these metric variables in the quadratic action and the definition \eqref{defdeltav} for $v_{\rm m}$, and writing explicitly only the terms that are quadratic in derivatives, neglecting those that are irrelevant in the kinetic limit, one finds the following action,
\be
\begin{split}
\label{PiKinet1}
S_{\rm kinetic}=&\int  d^4x a^3 M^2 \bigg\{  - 3 \dot \Psi_E^2  +\frac{1+\alphaT}{a^2} \big[ (\nabla\Psi_E)^2 -2\nabla\Phi_E\nabla\Psi_E \big]  \\
&+\frac{   \alpha \, H^2 }{2 (1+\alphaH)^2} \bigg[\left(1+\frac{c_s^2 }{c_{\rm m}^2 } \lambda^2 \right)\, \dot \pi^2 -  c_{s}^2  \frac{(\gpi)^2}{a^2}\bigg]  \\
&+\frac{\rho_{\rm m}+p_{\rm m}}{2 c_{\rm m}^2  M^2}\bigg[{\dot v_{\rm m}}^2- c_{\rm m}^2\frac{{(\nabla v_{\rm m} )}^2}{a^2} +\frac{2 \alphaH }{1+\alphaH  } \; \dot v_{\rm m}\,\dot {\pi} \bigg]  \bigg\}\,,
\end{split}
\ee
where $\lambda^2$ is the parameter encoding KMM, defined in eq.~\eqref{gpar}. Since here we are using the Jordan frame metric, where $\alphaDm= \alphaXm=0$, its definition reads
\be
\lambda^2 = \frac{3 }{ \alpha c_s^2} \alphaH^2 (1+w_{\rm m}) \Omega_\tm\;,
\ee
so that $\lambda$ is proportional to $\alphaH$.
Notice in the third line the presence of a kinetic coupling between the scalar and matter fields, $ \dot v_{\rm m} \dot \pi $, proportional to $\alphaH$.

Moreover,
at this order in derivatives the dynamics of $\pi$ and $v_{\rm m}$  is decoupled from that of  $\Phi_E$ and $\Psi_E$ and we can study them separately.
To simplify the analysis, we introduce the canonically normalized fields
\be
\pi_{\rm c}\equiv \frac{H M \alpha^{1/2} }{1+\alphaH}\, \pi\, , \qquad v_{\rm c}\equiv \left(\frac{\rho_{\rm m} +p_{\rm m}}{c_{\rm m}^2 }\right)^{1/2} \, v_{\rm m} \; ,
\ee
and we neglect the expansion of the universe, which is irrelevant in the kinetic limit. Then the dynamics is described by the Lagrangian
\be
\label{PiKinet2}
{\cal L} = \frac{1}{2}  \bigg\{  \bigg(1+ \frac{c_s^2}{c_{\rm m}^2} \lambda^2 \bigg) \dot \pi_{\rm c}^2 -  c_{s}^2  {(\gpi_{\rm c})^2} + {\dot{v}_{\rm c}}^2- c_{\rm m}^2 {{(\nabla v_{\rm c})}^2}  + 2 \frac{c_s}{c_{\rm m}} \lambda  \; \dot{v}_{\rm c}\,\dot {\pi}_{\rm c}\bigg\}\,.
\ee
In Fourier space, this gives the coupled system of equations
\be
\frac{d^2}{dt^2}  \begin{pmatrix}
\pi_c \\
v_c
\end{pmatrix}  + k^2 \begin{pmatrix}
c_{s}^2 & - \lambda \, c_s \, c_{\rm m}\\
- {\lambda \, c_{s}^3}/{c_{\rm m}}   &  c_{\rm m}^2 +\lambda^2 \, c_s^2
\end{pmatrix} \begin{pmatrix}
\pi_c \\
v_c
\end{pmatrix} =0 \;,
\ee
with normal modes
\be
\begin{pmatrix}
 c_{s}^3 \lambda/c_{\rm m}  &  c_{-}^2-c_{s}^2 \\
- c_s^3 \lambda /c_{\rm m} & c_s^2 - c_+^2
\end{pmatrix} \begin{pmatrix}
\pi_c \\
v_c
\end{pmatrix} \;,
\ee
where $c^2_\pm$ are the eigenvalues of the system, given by eq.~\eqref{cpm}.

As an example relevant for late-time cosmology, we consider the case where matter is described by a non-relativistic fluid (for instance CDM) with $w_\tm =0$ and $c_{\rm m}^2 =0$. Going back to standard normalization before setting $c_{\rm m}^2 =0$, the eigenmodes and respective eigenvalues of the system are
\begin{align}
\label{eigenm}
X_-= \ &v_\tm+\pi \,\frac{\alphaH}{1+\alphaH}\; , \qquad c_-^2=  c_{\rm m}^2 =0  \;, \\
X_+ = \ &\pi - v_\tm \, \lambda^2 \frac{1+\alphaH}{ \alphaH} \;, \qquad c_+^2=c_s^2 (1+\lambda^2)  \; , \label{Xp}
\end{align}
with $\lambda^2 =  3 \alphaH^2  \Omega_\tm / ({ \alpha c_s^2}) $.
While $X_+$ displays oscillations with frequency $\omega = \pm i c_+ k$, the speed of the fluctuations of $X_-$ vanishes as that of matter.

\subsection{Quasi-static regime}
\label{sec:QS}

Here we stick to the case where matter is non-relativistic, i.e.~$p_{\rm m} = c_{\rm m} = 0$, which applies to matter in late-time cosmology. When including the Hubble expansion, we expect  the oscillations of $X_+$ to get damped \cite{Sawicki:2015zya}.
In the absence of the oscillatory mode $X_+$, the time evolution is dominated by the Hubble friction and  time derivatives are of the order of the Hubble rate $H$. This is the quasi-static regime. We leave for the App.~\ref{app:QS} the discussion of how this regime is reached in the cosmological evolution.

In this case, focussing on the short-scale limit $k \gg k_+$, where $k_+$ denotes the sound horizon scale of the oscillating mode,
\be
\label{ks}
k_+ \equiv \frac{aH}{c_+} =  \frac{a H}{c_s \sqrt{1+\lambda^2}} \;,
\ee
and neglecting oscillations, the second-order action in Newtonian gauge becomes\footnote{To get the last term one can replace in eq.~\eqref{fullacmat}  the time derivative  of the field fluctuation $\delta \dot \sigma$ by its density fluctuation $ \delta \rho_{\rm m}$, using the expression
\be
\delta \rho_{\rm m} = \left(1 + \frac{1}{c_{\rm m}^2} \right) (\rho_{\rm m} + p_{\rm m}) \left( \frac{\delta \dot \sigma}{\dot \sigma_0} - \Phi \right) \;
\ee
valid for finite $c_{\rm m}^2$, and subsequently set $c_{\rm m}^2 =0$.}
\be
\begin{split}
\label{PiKinet1QS}
S =\int & d^4x a^3 M^2 \bigg\{\frac{1+\alphaT}{a^2} \big[ (\nabla\Psi_E)^2 -2\nabla\Phi_E\nabla\Psi_E \big]    - \frac{   \alpha H^2 c_{s}^2 }{2 (1+\alphaH)^2}   \frac{(\gpi)^2}{a^2}    - \Phi \frac{\delta \rho_\tm}{M^2} \bigg\}\,.
\end{split}
\ee
Variation of the above action with respect to $\Phi_E$ yields a Poisson-like equation for $\Psi_E$,
\be
 \frac{\nabla^2 \Psi_E }{a^2}  =   \frac32 H^2 \Omega_{\rm m}  \frac{ \delta_\tm}{1+\alphaH} \;. \label{PoissonEin}
\ee
In the above limit, also the scalar field fluctuations $\pi$ satisfy a Poisson-like equation. To derive it, one can vary the action \eqref{PiKinet1QS} with respect to $\pi$, taking into account that $\Phi_E$ and $\Psi_E$  depend on $\pi$ through the expressions \eqref{toE}.
Using eq.~\eqref{PoissonEin}, $\Phi_E = \Psi_E$, the definition of $\deltam$, eq.~\eqref{defdeltav}, and the continuity equation for matter,
\be
\deltamd = - \frac{\nabla^2 v_\tm}{a^2} \label{conti} \;,
\ee
one obtains
\be
 \frac{\nabla^2 \pi }{a^2}   = \frac{3 H \Omega_\tm}{c_s^2 \alpha} \left[  \left(  \xi - \frac{\dot \alpha_{\rm H} }{H}\right) \deltam  + \frac{\alphaH(1+\alphaH)}{H} \frac{\nabla^2 v_{\rm m}}{a^2}   \right] \;, \label{Poissonpi}
\ee
where we remind the reader that $\xi \equiv \alphaB (1+\alphaT) + \alphaT - \alphaM - \alphaH (1+\alphaM)$ (see eq.~\eqref{xidef}). Notice the presence of the last term on the right-hand side, proportional to the matter velocity, which stems from the KMM.
Indeed, by using the definition of the ``$+$'' eigenmode $X_+$, eq.~\eqref{Xp}, this equation can be rewritten as
\be
 \frac{\nabla^2 X_+ }{a^2}   = \frac{3 H \Omega_\tm}{c_s^2 \alpha}   \left(  \xi - \frac{\dot \alpha_{\rm H} }{H}\right) \deltam  \;,
\ee
which shows that after the oscillating regime ends, $X_+$ (and not $\pi$) satisfies a Poisson-like constraint equation.

Let us now derive the constraint equations for $\Psi$ and $\Phi$.
We can  rewrite equation \eqref{PoissonEin} in terms of $\Psi$ using the definition of $\Psi_E$, eq.~\eqref{toE}, and eq.~\eqref{Poissonpi}.
We can then use $\Phi_E = \Psi_E$ and solve eqs.~\eqref{PoissonEin}, \eqref{Poissonpi} and its derivative to find an equation for $\Phi$.
This yields
\begin{align}
\label{nablapsi}
  \frac{\nabla^2 \Psi}{a^2}
&= \frac{3}{2}  H^2 \Om_{\rm m}    \mu_{\Psi} \delta_{\rm m} +  \lambda^2 \left( \frac{\alphaB }{\alphaH } - 1 \right)  H \frac{\nabla^2 v_{\rm m}}{a^2}    \;, \\
\label{nablaphi}
  \frac{\nabla^2 \Phi}{a^2}
&=  \frac32 H^2 \Omega_\tm \mu_\Phi \deltam + \gamma  H \frac{\nabla^2 v_{\rm m}}{a^2 } \;,
\end{align}
where $\mu_{\Psi}$ and $\mu_\Phi$ are defined as
\begin{align}
\label{muphi}
\mu_\Psi & \equiv \frac{1}{ 1+ \alphaH} \bigg[    1+ \frac{2(\alphaB - \alphaH)}{c_s^2 \alpha}   \left(\xi - \frac{\dot \alpha_{\rm H}}{H} \right)
  \bigg] \;, \\
\label{Geff}
\mu_\Phi &\equiv  \frac{1}{(1+  \lambda^2 ) (1+ \alphaH)^2} \left\{ c_T^2 + \frac{2\xi}{c_s^2 \alpha} \left( \xi - \frac{ \dot \alpha_{\rm H} }{H} \right) +
a \alphaH (1+\alphaH) \left[ \frac{2 }{a H c_s^2 \alpha } \left(\xi - \frac{ \dot \alpha_{\rm H} }{H} \right)  \right]^{\hbox{$\cdot$}}  \right\} \;,
\end{align}
and $\gamma$ is defined as
\begin{align}
\label{gamma}
\gamma  \equiv  \frac{d \ln \left(1+ \lambda^2 \right)}{d \ln a}  \;.
\end{align}
The parameters $\mu_\Psi$ and $\mu_\Phi$ represent modifications of the Poisson law, respectively for $\Psi$ and $\Phi$,  and are equal to one in the standard case.
The last term on the right-hand side of eqs.~\eqref{nablapsi} and \eqref{nablaphi} proportional to the Laplacian of the  matter velocity
potential
vanishes in the absence of  KMM.

Equation \eqref{nablaphi}, together with the continuity equation \eqref{conti} and the Euler equation,
\be
\dot v_\tm = - \Phi \label{Euler}\;,
\ee
can be used to derive a closed second-order equation for the matter density contrast in the quasi-static limit. Indeed, taking the time derivative of the continuity equation, and
plugging in the latter the Euler equation and eq.~\eqref{nablaphi}, one obtains \cite{Kobayashi:2014ida},
\be
\label{deltaevol}
\deltamdd + (2 + \gamma) H \deltamd =  \frac32 H^2 \Omega_\tm \mu_\Phi \deltam   \;.
\ee

A comment on this equation is in order here. For $\alphaH=0$,
the friction term  vanishes, $\gamma=0$,
and the strength of gravitational clustering is modified by  \cite{Gleyzes:2015rua}
\be
\mu_\Phi  =   c_T^2 + \frac{2 \xi^2}{c_s^2 \alpha}   \;, \qquad (\alphaH=0) \;, \label{muPhiHorn}
\ee
which, for $c_T^2 \geq 1$,\footnote{Cosmic rays observations put tight constraints on a propagation speed $c_T^2< 1$~\cite{Moore:2001bv}. Another lower bound can be put from  binary pulsar orbital periods \cite{Jimenez:2015bwa}.} is always larger than one. Thus, the exchange of the fifth force tends to enhance gravity on small scales \cite{Piazza:2013pua,Gleyzes:2015pma,Gleyzes:2015rua,Pogosian:2016pwr}. On the contrary, in the presence of KMM $\mu_\Phi - c_T^2$ can be negative, corresponding to a repulsive scalar fifth-force, thus weakening gravity. Moreover, the last term on the right-hand side of \eqref{nablaphi} can act as a friction term for structure formation. This results in a suppression of clustering, even for a $\Lambda$CDM background evolution. We will see an explicit example below.

For completeness and comparison with observations, we provide here also the expression of the Weyl
potential, obtained by summing eqs.~\eqref{nablapsi} and \eqref{nablaphi},
\be
\label{Weyl}
\begin{split}
 &  \frac{1}{a^2 H^2 } \nabla^2 ( \Phi+\Psi ) =  \frac32 \Omega_\tm ( \mu_\Psi  + \mu_\Phi   ) \deltam +  \left[  \left( 1 -\frac{\alphaB}{\alphaH} \right) \lambda^2 -{\gamma}   \right] \frac{\dot \delta_{\rm m}}{ H }  \;,
\end{split}
\ee
where we have used the continuity equation to replace the velocity $v_{\rm m}$ by $\dot \delta_{\rm m}$.
Note that for $\alphaH =0 $, the equations in this section reduce to their analogous expressions derived for instance in \cite{Gleyzes:2015rua}.

\section{Observational signatures of Kinetic Matter Mixing}
\label{sec:alphaH}

In this section we discuss the effects of KMM on the  power spectrum of the  matter density contrast and on the CMB.
In particular, we compute the comoving matter density contrast, defined as
\be
\Delta_{\rm m} \equiv \delta_{\rm m} - 3 H v_{\rm m} \;,
\label{comovingdensity}
\ee
where $\deltam$ and $v_{\rm m}$ are in Newtonian gauge.
For the CMB we focus on the  lensing potential and the temperature fluctuations.

The observables are computed using COOP~\cite{zqhuang_2016_61166}, which solves  linear perturbations in Newtonian gauge and in the Jordan frame, assuming minimal coupling of all matter species.  In the $\Lambda$CDM case, COOP evolves $\Psi$, $\Psi_{N_{e}}\equiv d\Psi/dN_{e}$ and matter perturbations, where $N_{e} \equiv \ln a$ is the program time variable. The detailed algorithm and equations can be found in Ref.~\cite{Huang:2012mt}. To describe deviations from $\Lambda$CDM using the Effective Theory of Dark Energy, COOP evolves two additional variables, $\mu \equiv H \pi$  and $\mu_{N_{e}}\equiv d\mu/dN_{e} $. In the Jordan frame, only the metric perturbations are coupled to $\mu$ and $\mu_{N_{e}}$. The evolution equations of $\Psi_{N_{e}}$ and $\mu_{N_{e}}$ are obtained by eliminating $\Phi$ from eqs.~(111)--(113) in Ref.~\cite{Gleyzes:2014rba}. For numeric stability, COOP
combines the energy conservation equation and the pressure equation, respectively eqs.~(109) and (112) of Ref.~\cite{Gleyzes:2014rba},
such that the evolution equation of $\Psi_{N_{e}}$ has a traceless source, i.e.~it is of the form
$d\Psi_{N_{e}}/dN_{e} = \ldots + (\delta p_{\rm m} - \frac{1}{3}\delta \rho_{\rm m})/(2M^2)$.
See Ref.~\cite{Huang:2012mt} for more details
on
this technique. Once the linear perturbations are solved, COOP computes CMB power spectra using a line-of-sight integral~\cite{Seljak:1996is, Hu:1997hp}. Matter power spectra are computed via a gauge transformation from the Newtonian to the CDM rest-frame synchronous gauge.

For the cosmological parameters we use the Planck TT+lowP parameters \cite{Ade:2015xua}. In particular, we assume a physical density of baryons and CDM respectively given by $\Omega_{b,0} h^2 = 0.02222$ and $\Omega_{c,0} h^2 = 0.1197$, we fix the acoustic scale at recombination as $\theta = 1.04085 \times 10^{-2}$, the amplitude of scalar primordial fluctuations $A_s = 2.2 \times 10^{-9}$, the scalar spectral tilt $n_s = 0.9655$ and the reionization optical depth $\tau = 0.078$. We assume that the background expansion history is the same as in $\Lambda$CDM. This implies that $h = 67.31$ and $\Omega_{\rm m,0} = 0.315$.
Initial conditions are taken to be adiabatic (see e.g.~\cite{Gleyzes:2014rba}).

To focus on the effects of KMM, we set
\be
\alphaB= \alphaM=\alphaT=0 \;.
\ee
Moreover, we parametrize the time dependence of
$\alphaK$ and $\alphaH$  as
\be
\alphaK =  \alpha_{\rm K,0} \frac{\Omega_{\rm DE} (t) }{\Omega_{\rm DE,0}}\;, \qquad
\alphaH =  \alpha_{\rm H,0} \frac{\Omega_{\rm DE} (t) }{\Omega_{\rm DE,0}}\;,
\ee
where $ \Omega_{\rm DE}$ is the fractional energy density of dark energy, defined as $ \Omega_{\rm DE}  \equiv 1 - \sum_I \Omega_I $, where the sum is over all matter species (baryons, photons, neutrinos and
CDM).

For the sake of clarity, in the following discussion we will simplify the above parametrization and consider only baryons and
CDM
in the matter sector. This is justified by the fact that according to this parametrization, the effects of dark energy become relevant only at late time.
However,  we stress that the numerical calculation performed with COOP contains the full matter sector, including (massless) neutrinos.
Under these simplifying assumptions the background expansion history becomes
\be
H^2  = H_0^2 \left[ \Omega_{\rm m 0} a^{-3} +1-\Omega_{\rm m 0}  \right] \;.  \label{Hparametr}
\ee
Moreover, in this case the speed of scalar fluctuations (see eq.~\eqref{tildecss}) simplifies to
\be
\label{csqH}
c_s^2 = \frac{\alphaH  \big[ 2 + 3 \Omega_{\rm m} (1- \alphaH) \big]}{\alphaK} \;.
\ee
Requiring the absence of ghosts ($\alpha> 0$, see definition in eq.~\eqref{alpha_def}) and gradient instabilities, respectively implies that
\be \label{stabWin}
\alphaK \ge 0\,, \qquad 0 \le \alphaH \le 1 +\frac{2}{3 \Omega_{\rm m}}\; .
\ee
In the following we set the current value of $\alphaK$  to unity, $\alpha_{\rm K,0} =1$ and we plot the effect of $\alphaH$ in terms of  four different values of this parameter today, i.e.~$\alpha_{\rm H,0} = 0.06$, $0.12$, $0.24$ and $0.48$, which
are
always in the stability window~\eqref{stabWin}. Note that to avoid that scalar fluctuations become superluminal in the past we must  require
\be
\alpha_{\rm H} \le \frac15 \alpha_{\rm K} \; . \label{superlum}
\ee
Just for the purpose of illustration, in the next two subsections we ignore constraints from superluminality, as we need large values of $\alphaH$ to better visualise the effects on the observables.

\subsection{Matter power spectrum}
\label{sec:mps}

\begin{figure}[h]
\centerline{\includegraphics[width=0.504\textwidth]{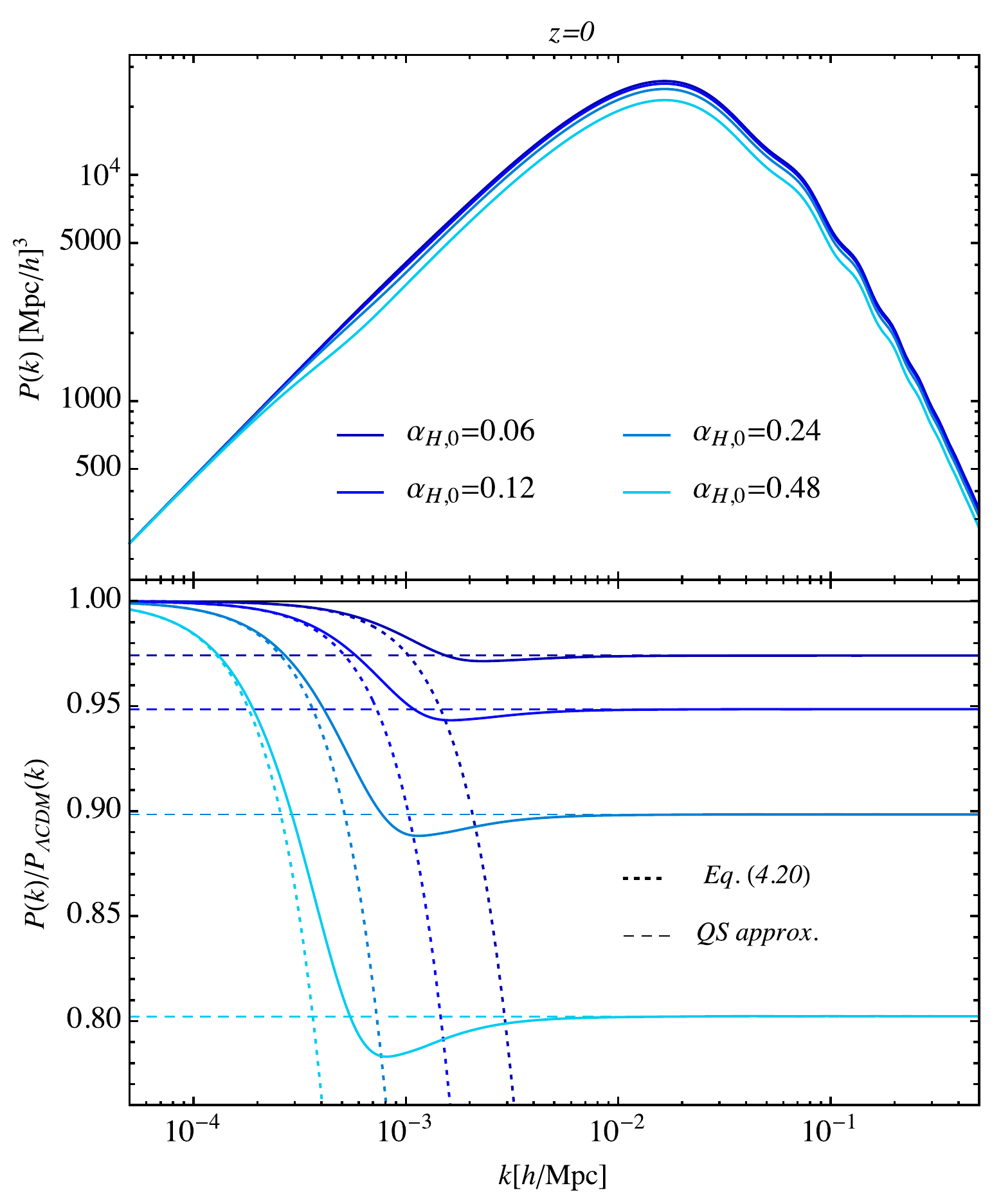}
\includegraphics[width=0.5\textwidth]{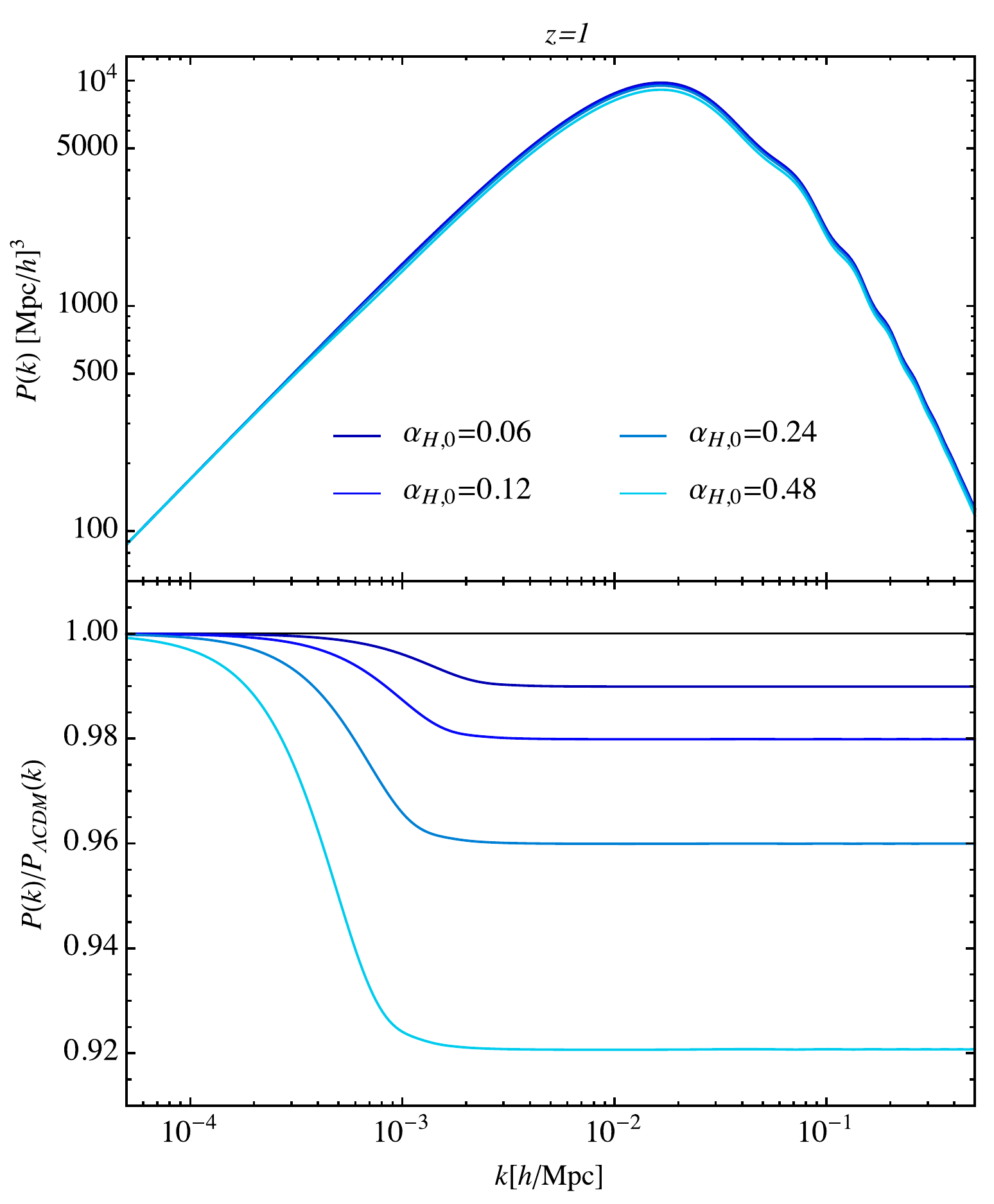}}
\caption{Effect of KMM on the matter power spectrum for four different values of $\alphaH$ today, i.e.~$\alpha_{\rm H,0} = 0.06$, $0.12$, $0.24$ and $0.48$, at redshift $z=0$ (left panel) and $z=1$ (right panel). The lower plots display the ratio of these power spectra with the respective spectra for $\alphaH=0$. For comparison, the dashed and dotted lines in the left lower panel respectively show the quasi-static approximation and the perturbative solution of eq.~\eqref{Deltam2nd}.}
\label{fig:alphaH_PS}
\end{figure}
On short scales, increasing $\alpha_{\rm H,0}$ suppresses the power spectrum of matter fluctuations, shown as a function of $k$ in Fig.~\ref{fig:alphaH_PS}.
On these scales we can neglect the velocity potential in the  definition
of the comoving matter density contrast, eq.~\eqref{comovingdensity}, which reduces to $\delta_{\rm m}$ in the Newtonian gauge, $\Delta_{\rm m} \approx \delta_{\rm m}$.
Moreover, to understand the power suppression we can apply the quasi-static approximation, i.e.~eq.~\eqref{deltaevol}. Specializing to the case with only nonvanishing $\alphaK$ and $\alphaH$ and using the time parametrization above, $\mu_\Phi $ and $\gamma$ in eq.~\eqref{deltaevol}, defined in eqs.~\eqref{Geff} and \eqref{gamma}, become
\begin{figure}[t]
\centerline{\includegraphics[width=0.65\textwidth]{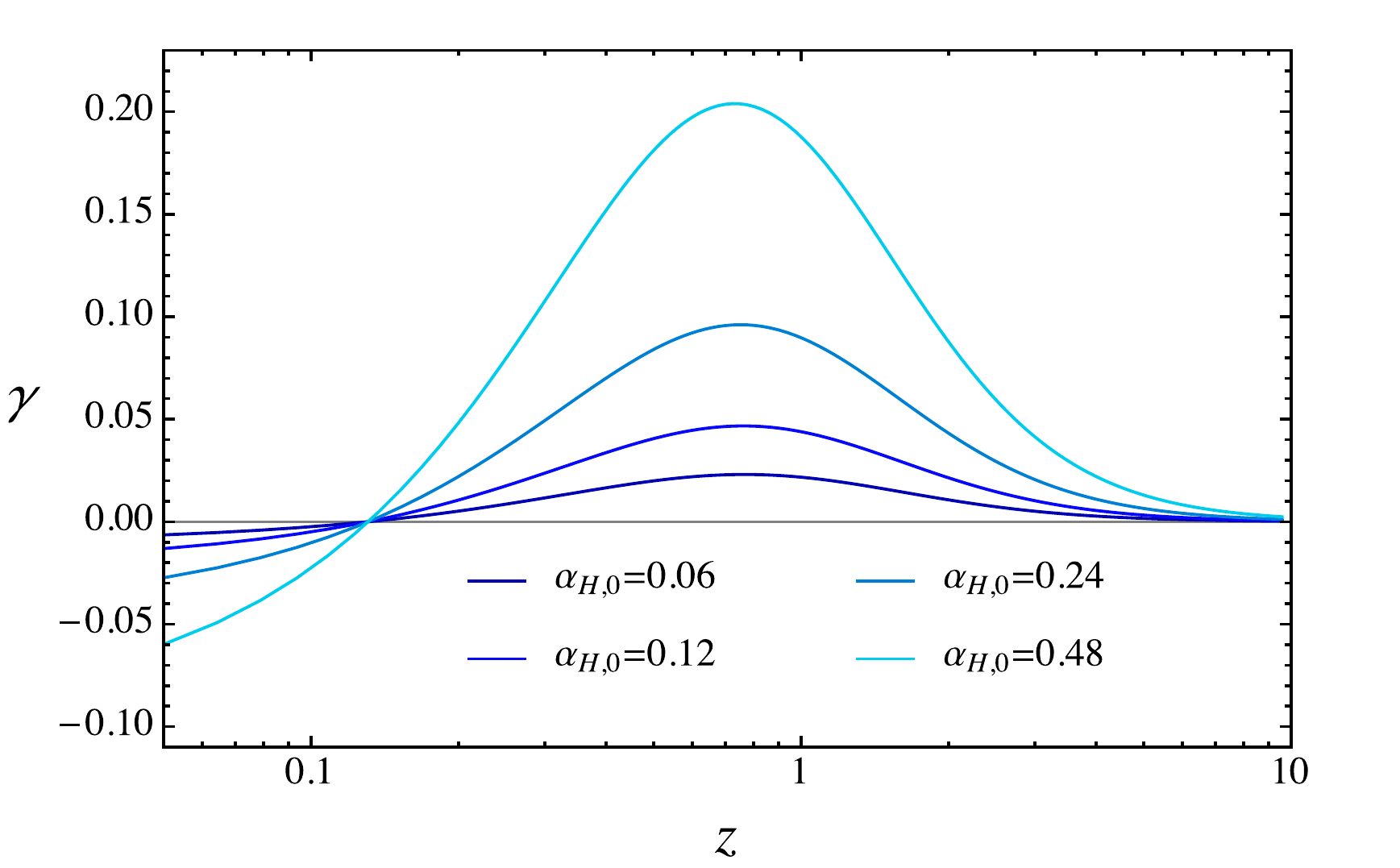}}
\caption{Friction term $\gamma$ given in eq.~\eqref{gammaGeff}, as a function of redshift.}
\label{fig:gamma}
\end{figure}
\be
\mu_\Phi = 1 - \gamma \;, \qquad  \gamma = - \Omega_{\rm m} \frac{9  \alphaH  (2 - 4 \Omega_{\rm m} - 3 \Omega_{\rm m}^2 )}{(2+3 \Omega_{\rm m}) \big[ 2 +3 (1-\alphaH) \Omega_{\rm m}\big] } \;. \label{gammaGeff}
\ee
The friction term $\gamma$ as a function of redshift  is plotted   in Fig.~\ref{fig:gamma}. It starts positive and changes sign only recently, when $\Omega_{\rm m} = (\sqrt{10} -2 )/3 \simeq 0.39$. In particular, during matter domination it behaves as
\be
\gamma = \frac95 \alphaH  +{\cal O} (\Omega_{\rm DE}^2)  \;,
\ee
where we have  expanded in $\Omega_{\rm DE}$.
Thus, $\gamma$ suppresses the power spectrum with respect to the $\Lambda$CDM case and the effect is linear in $\alphaH$. The modification of the Poisson equation has an analogous effect:  $\mu_\Phi$ starts smaller than unity decreasing the strength of gravity, and gets larger than one only when $\gamma$ changes sign. This has again the cumulative effect of suppressing the  power spectrum  with respect to the $\Lambda$CDM case. We have checked that eq.~\eqref{deltaevol}, with $\gamma$ and $\mu_\Phi$ given above, reproduces the suppression observed in Fig.~\ref{fig:alphaH_PS}.

Corrections to the quasi-static approximation  are expected to be of the order ${\cal O} ( k_+^2 / k^2 )$, where $k_+ $ is the sound horizon scale defined in eq.~\eqref{ks}.\footnote{The sound horizon
scale
is $k_+ \simeq 8.1$, $5.8$, $4.2$ and $3.1 \times 10^{-4} h/$Mpc at redshift $z=0$  and $k_+ \simeq 5.9$, $4.2$, $3.0$ and $2.2 \times 10^{-4} h/$Mpc at redshift $z=1$, respectively for $\alpha_{\rm H,0} = 0.06$, $0.12$, $0.24$ and $0.48$.}
Thus, on larger scales this approximation fails to reproduce the correct spectrum, as shown in the figure. However, we can find an integral solution for the density perturbation on the largest scales by solving the Einstein and scalar field equations perturbatively in $\alphaH$ (while keeping the exact dependence on $\alphaK$ to avoid inconsistencies \cite{Iglesias:2007nv}).
For the parametrization chosen in this section, these equations read
\begin{align}
 - \frac{k^2}{a^2} \Psi + \dot H  (\Delta_{\rm m} + \Delta_{\rm DE}) & = 0 \label{E1} \;,\\
\Phi - \Psi & = \alphaH (\dot \pi - \Phi) \;, \label{E3} \\
\ddot \Psi + H (3 \dot \Psi + \dot \Phi) + (2 \dot H + 3 H^2) \Phi & =0 \;, \label{E4} \\
 \dot \Delta_{\rm DE}  &=  - \alphaH \frac{k^2}{a^2} \frac{H}{\dot H} (\dot \pi -\Phi) \;, \label{Epi}
\end{align}
where we have used the background Friedmann equations and the comoving energy density contrast associated to dark energy, $\Delta_{\rm DE}$, is defined as
\be
\label{Dde}
\Delta_{\rm DE} \equiv - \frac{\alphaK}{2} H^2 (\dot \pi -\Phi) - \alphaH \frac{k^2}{a^2} (\Psi + H \pi) \;.
\ee
Equation~\eqref{E1} has been obtained from combining the ``$00$'' and ``$0i$'' scalar components of the Einstein equations, eqs.~\eqref{E3} and \eqref{E4} are respectively the traceless and trace part of the ``$ij$'' scalar components of the Einstein equations and eq.~\eqref{Epi} is the evolution equation of $\pi$.\footnote{The complete Einstein equations can be found in \cite{Gleyzes:2014rba}.}
(The evolution equations for the matter density contrast $\Delta_{\rm m}$ is automatically satisfied by these equations.)

In the absence of KMM, i.e.~for $\alphaH=0$,  eqs.~\eqref{E1}--\eqref{Epi} are solved by the standard $\Lambda$CDM solution with adiabatic initial conditions \cite{Gleyzes:2014rba}, i.e.
\be
\Phi = - \dot \epsilon\;, \qquad \Psi = H\epsilon - \zeta_0\;,   \qquad \Delta_{\rm m} = \Delta_{{\rm m}}^{(0)} \equiv \frac{k^2}{a^2} \frac{H \epsilon - \zeta_0}{\dot H} \;, \qquad \Delta_{\rm DE} =0 \;,  \qquad \left( \alphaH = 0   \right)  \label{solalphaHzero}
\ee
where $\zeta_0$ is the (conserved) comoving curvature perturbation on super-Hubble scales and $\epsilon$ is defined as
\be
\epsilon  \equiv \frac{\zeta_0}{a} \int a dt \;.
\ee
Notice that, for $\alphaH=0$, $\pi = -\epsilon$
and thus the combination $\dot \pi - \Phi $ vanishes. Hence, the right-hand sides of eqs.~\eqref{E3} and \eqref{Epi} vanish also at first order  in $\alphaH$ and eq.~\eqref{solalphaHzero} keeps being a solution of  the above equations.

The combination $\dot \pi - \Phi $ does not vanish at first order in $\alphaH$.  Using in eq.~\eqref{Dde} that $\Delta_{\rm DE}=0$ at this order, one obtains
\be
\dot \pi - \Phi = 2 \zeta_0 \frac{k^2}{a^2H^2}  \frac{\alphaH}{\alphaK} +{\cal O}(\alphaH^2) \;.
\ee
Thus, deviations from $\Lambda$CDM arise  at second-order in $\alphaH$, as the backreaction
effect
of $\pi$ on gravity. This is similar to what happens in the context of the Ghost Condensate, where the mixing of the  scalar fluctuations with gravity gives rise to a Jeans-like instability also on a $\Lambda$CDM background \cite{ArkaniHamed:2003uy,Creminelli:2008wc}.
It is now straightforward to find the solution for $\Delta_{\rm m} $ at this order in $\alphaH$, by replacing the second-order solution for $\Psi$ and $\Delta_{\rm DE}$ in eq.~\eqref{E1}. The former can be derived by solving eq.~\eqref{E4} after replacing $\Phi$ from eq.~\eqref{E3}. This yields
\be
\Psi= H\epsilon - \zeta_0 + 2 \zeta_0 \frac{\alphaH^2}{\alphaK} \frac{k^2}{a^2 H^2 } \left( 1 - aH^5  \int \frac{dt }{aH^4} \right) +{\cal O}(\alphaH^3)\;.
\ee
The latter can be derived from eq.~\eqref{Epi},
which yields
\be
\Delta_{\rm DE} = -2 \zeta_0 \frac{\alphaH^2 }{\alphaK} \frac{k^4}{a^4} \frac{a H^2 }{\dot H} \int \frac{dt}{aH^3}  +  {\cal O} (\alphaH^3) \;.
\ee
Thus, one obtains
\be
\Delta_{\rm m} =  \Delta_{{\rm m}}^{(0)} \left[ 1 - 2 a H^2  \frac{\alphaH^2}{\alphaK} \frac{k^2}{a^2}  \left( \int \frac{dt}{a H^3} - H \int \frac{dt }{aH^4} \right) \left(1 - \frac{H}{a} \int a dt  \right)^{-1}  +  {\cal O} (\alphaH^3) \right] \;. \label{Deltam2nd}
\ee
Notice that this solution breaks down on small scales because the quasi-static limit assumes $\alphaH \neq 0$.

On very large scales, i.e.~for
\be
\begin{split}
k \lesssim k_* & \equiv  \frac{\sqrt{\alphaK}}{ \sqrt{2} \, \alphaH}\frac{a}{H}   \left(1- \frac{H}{a} \int a dt \right)^{1/2 }  \left(  a \int \frac{dt}{a H^3} -   {Ha} \int \frac{dt }{aH^4} \right)^{-1/2} \\
& \simeq  \frac{\sqrt{\alpha_{\rm K,0}}}{\alpha_{\rm H,0}} \times 5.4 \times 10^{-4} h /\text{Mpc} \;,
\end{split}
\ee
the power spectrum is unmodified by KMM, although this restricts only to the case where the background expansion is that of $\Lambda$CDM. On intermediate scales, $k_* \lesssim k \lesssim k_+ $, the power spectrum drops as $k^2$ due to the second term on the right-hand side of eq.~\eqref{Deltam2nd}.

\subsection{Cosmic Microwave Background}

\begin{figure}[h]
\centerline{\includegraphics[width=0.525\textwidth]{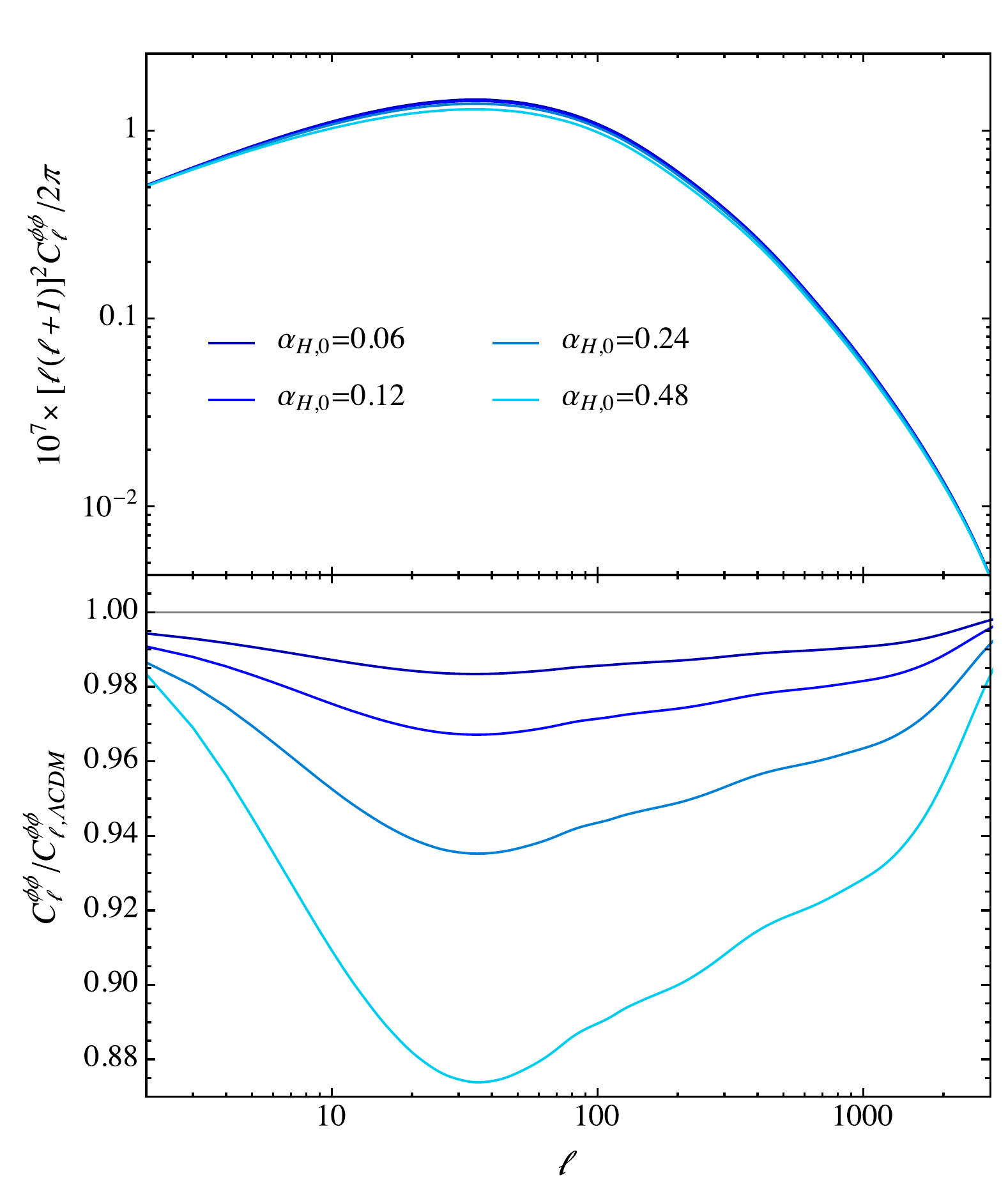}
\includegraphics[width=0.52\textwidth]{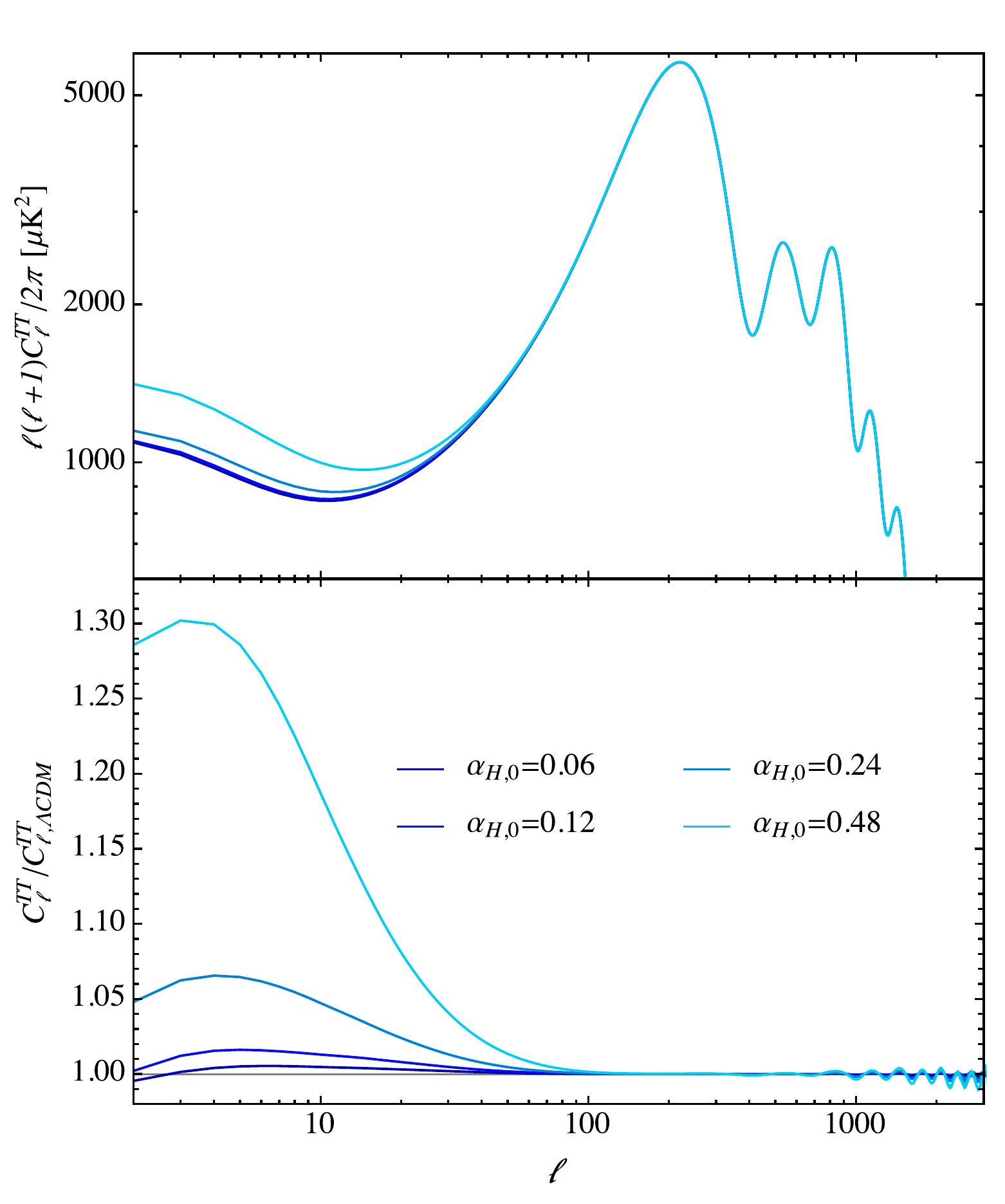}}
\caption{Effect of KMM ($\alphaH$) on the CMB lensing potential  (left panel) and on the CMB anisotropies (right panel) angular power spectra. The lower plots display the ratio of these angular spectra with the respective spectra for $\alphaH=0$.}
\label{fig:alphaH_CMB}
\end{figure}
In Fig.~\ref{fig:alphaH_CMB}, on the left panel we plot the angular power spectrum of the CMB lensing potential,
defined as \cite{Lewis:2006fu}\footnote{This is not to be confused with the scalar field $\phi$ introduced in Sec.~\ref{section2}.}
\be
\phi (\hat n) = -\int_0^{z_*} \frac{d z}{H(z)}  \frac{\chi (z_*) - \chi (z)}{\chi (z_*)  \chi (z) } \big[ \Phi (\chi  \hat n, z) + \Psi (\chi  \hat n, z) \big] \;,
\ee
where $\chi \equiv \int_0^z dz/H(z)$ is the conformal distance and $z_*$ denotes the redshift of last scattering.
On the right panel, we plot the angular power spectrum of the CMB anisotropies as a function of the multipole $l$.
As a rough approximation, we can understand the CMB lensing potential by looking at the Weyl potential $(\Phi + \Psi) / 2$ in the quasi-static regime, i.e.~using eq.~\eqref{Weyl}. Indeed, the bulk of the CMB lensing kernel is at $ 0. 5\lesssim z  \lesssim 6$ \cite{Lewis:2006fu}, where deviations from  this approximation are below $\sim 5\%$ for the values of $\alpha_{\rm H,0}$ that we considered.

Let us define the  quantity \cite{Gleyzes:2015pma}
\be
\label{lensingmu}
\mu_\text{\rm WL} \equiv  \frac{2 \nabla^2 ( \Phi+\Psi )}{3 a^2 H^2 \Omega_\tm \delta_\tm}  \;.
\ee
For $\Lambda$CDM, $\mu_\text{\rm WL}= 2$; in general,
this quantity
characterizes the deviations in weak lensing observables  from the $\Lambda$CDM case.
This definition cannot be directly applied to eq.~\eqref{Weyl}, because of the presence of the terms proportional to $\dot \delta_{\rm m}$ on the right-hand side of this equation. In the presence of KMM, $\alphaH \neq 0$, these terms equally contribute to the modifications of the Weyl potential as those proportional to $\delta_\tm$ and cannot be neglected. However,  a fair approximation to simplify the discussion is to replace $\dot \delta_{\rm m}$ by its expression in matter domination,   $\dot \delta_{\rm m} \simeq H   \delta_{\rm m}$.
Setting $\alphaB=\alphaM=\alphaT=0$
and employing the approximation above in  eq.~\eqref{Weyl}, the effect of $\alphaH$ in weak lensing observables can be rewritten as
\be
\label{mulensingH}
\mu_\text{\rm WL} -    2  =   \alphaH \frac{8 - 9 \Omega_\tm (1+ \Omega_\tm)}{2 + 3(1-\alphaH) \Omega_\tm}  \;.
\ee
One can verify that this quantity is negative for $z \gtrsim 0.5$, i.e.~inside the bulk of the CMB lensing kernel. Therefore, the lensing potential is suppressed by the modification of gravity induced by $\alphaH$.
For small $\Omega_{\rm DE}$, in matter domination this suppression is roughly proportional to $\alphaH$, as observed in Fig.~\ref{fig:alphaH_CMB}. Expanding at linear order in $\Omega_{\rm DE}$, the above relation simplifies to $\mu_\text{\rm WL} -    2 =  - 2 \alphaH + {\cal O}(\Omega_{\rm DE}^2)$.

Let us now turn  to the CMB anisotropies, right panel of Fig.~\ref{fig:alphaH_CMB}. At large $l$, the anisotropies are completely unaffected by the KMM because they are  generated at recombination,\footnote{Because of this, polarization is also unaffected. For this reason we only show the temperature spectrum.} when $\alphaH$ vanishes. The only visible effect is an oscillating pattern observed at high $l$ (noticeable in the lower right panel of Fig.~\ref{fig:alphaH_CMB}.), due to the change in the CMB lensing discussed above. Indeed,  lensing smears the CMB acoustic peaks; for larger values of $\alpha_{\rm H,0}$ the smearing  is suppressed and CMB peaks  enhanced.

At low $l$, the deviations from the $\Lambda$CDM case are dominated by the ISW effect,
which depends on the time variation of the Weyl potential, i.e.
\be
\frac{\Delta T}{T}^{\rm ISW} (\hat n) = - \int_0^{z_*} dz  \big[  \partial_z \Phi  (\chi  \hat n, z) + \partial_z  \Psi (\chi  \hat n, z) \big] \;.
\ee
Taking the derivative of eq.~\eqref{lensingmu} with respect to the $e$-foldings,  one obtains the following relation, which only holds in the quasi-static limit:
\be
\left. \frac{d \ln (\Phi+\Psi) }{ d \ln a} \right|_{\rm QS}  =  f_{\rm QS} -1 + \frac{d \ln \mu_{\rm WL}}{d \ln a} \;,
\label{ISW}
\ee
where
\be
f_{\rm QS} \equiv \left. \frac{ d \ln \delta_\tm }{d \ln a} \right|_{\rm QS} \;,
\label{fQS}
\ee
is the growth rate computed using the quasi-static approximation.
In $\Lambda$CDM, $\mu_{\rm WL} = 2$ and the time variation of $\Phi+\Psi$ is given by the first two terms on the right-hand side, i.e.~the deviation of the matter growth rate from unity, which is negative. When gravity is modified, the last term on the right-hand side does not vanish. In the case of KMM, it  contributes with the same sign as the first term, enhancing the ISW effect. For example, assuming matter domination and expanding in $\alphaH$ one finds
\be
\frac{d \ln \mu_{\rm WL}}{d \ln a} =  - 3 \alphaH + {\cal O}(\Omega_{\rm DE}^2)\;,
\ee
which explains the enhancement in the ISW effect observed in the right panel of Fig.~\eqref{fig:alphaH_CMB}, roughly proportional to $\alphaH$.

\subsection{Short-scale tension}
\label{sec4.3}

An intriguing issue that recently came up is the tension between the overall normalization of density fluctuations on large scales, inferred from the CMB anisotropies, and the amplitude of density fluctuations on small scales, measured with the large scale structures  at low redshift. In particular, the value of $\sigma_8$---defined as the rms of the fractional density fluctuation in a sphere of $8h^{-1}$Mpc---computed from the weak lensing measurements of the Canada-France Hawaii Telescope Lensing Survey (CFHTLens)\cite{Kilbinger:2012qz,Heymans:2013fya, Kitching:2014dtq,Kohlinger:2015tza} and from cluster counts \cite{Ade:2013lmv,Ade:2015fva,deHaan:2016qvy} appears to be lower than the one inferred from CMB measurements by Planck~\cite{Ade:2013zuv,Ade:2015xua}. This tension has been recently confirmed by the tomographic weak gravitational lensing analysis of the Kilo Degree Survey (KiDS) \cite{Hildebrandt:2016iqg}, while it has been alleviated by the analysis of the latest data of the SDSS-III Baryon Oscillation Spectroscopic Survey~\cite{Alam:2016hwk}. Another aspect of  this tension is reflected in redshift space distortion measurements~\cite{Macaulay:2013swa}, which indicate that the combination of $f \sigma_8$---where $f \equiv d \ln \delta_{\rm m} / d \ln a$ is the growth factor---is lower with respect to the value inferred from the Planck results.

\begin{figure}[t]
\centering
\includegraphics[width=0.6\textwidth]{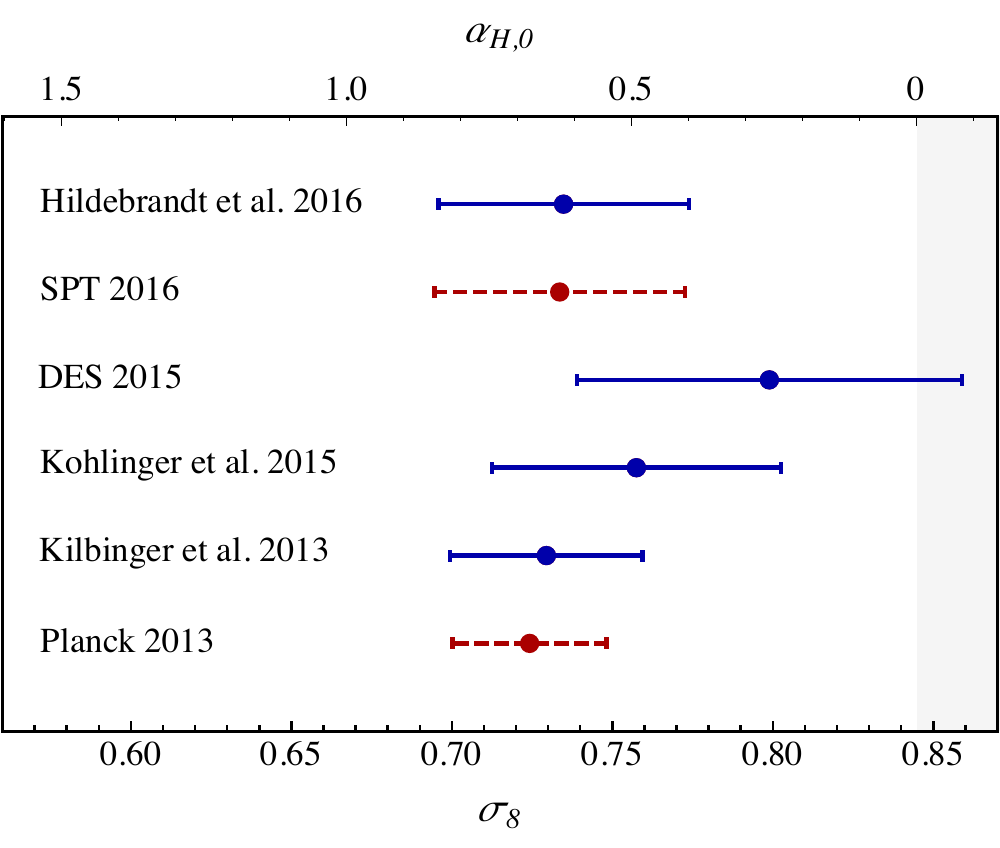}
\caption{Relation between $\alpha_{\rm H,0}$ and the corresponding $\sigma_8$ at redshift $z=0$, calculated using eq.~\eqref{sigma8}, respectively in the top and bottom $x$-axes. The $\alpha_{\rm H,0} = 0$ line
corresponds
to $\Lambda$CDM and the region $\alpha_{\rm H,0} < 0$ is shaded because it is out of the stability window~\eqref{stabWin}.
The plot also shows the measurements of $\sigma_8$ and their respective 1-$\sigma$ errors from several
collaborations.\protect\footnotemark~In particular, the constraints based on cluster counts (red dashed lines) are from
Planck 2013~\cite{Ade:2013lmv} and SPT 2016 \cite{deHaan:2016qvy}.
The constraints based on weak lensing observations (blue solid lines) are from several analysis of the CFHTLens, by Kilbinger et al.~2013 \cite{Kilbinger:2012qz}, K\"ohlinger et al.~2015 \cite{Kohlinger:2015tza} and Hildebrandt et al.~2016 \cite{Hildebrandt:2016iqg}, and from the cosmic shear study of DES 2015 \cite{Abbott:2015swa}.}
\label{fig:sigma8}
\end{figure}

\footnotetext{An analysis of the effects of systematics on the CFHTLenS data, not shown in Fig.~\ref{fig:sigma8}, has been carried out by Joudaki
et al. in~\cite{Joudaki:2016mvz}. Moreover, we show the Planck 2013 cluster-based constraint because the more recent analysis by the Planck collaboration~\cite{Ade:2015fva} did not release numerical values. However, the Planck 2015 results were found in agreement with the previous ones \cite{deHaan:2016qvy}.}

Even though the tension is not extremely significant and depends on the uncertainties of the modeling of the non-linear scales and, for the redshift-space distortion measurements, of the galaxy bias, it might indicate a  deviation from the concordance model. For instance, some  attempts have been made to solve this tension using  massive (active and sterile) neutrinos \cite{Wyman:2013lza,Battye:2014qga}. However, the most recent Planck analysis seems to disfavour this solution \cite{Ade:2015xua}.

Given that little is known of the clustering properties of dark energy, it is natural to try to explain this tension by considering a  model where deviations from the concordance one are  restricted only on short scales.
A recent proposal  in this direction has been undertaken in \cite{Kunz:2015oqa} by exploiting the so-called ``dark degeneracy'' between dark matter and dark energy \cite{Kunz:2007rk} and replacing part of the dark matter by a perfect-fluid clustering dark energy with sound speed of fluctuations smaller than unity (see for instance \cite{Creminelli:2009mu,Sefusatti:2011cm} for a phenomenological study of clustering dark energy in the zero sound-speed limit).

More generally, one could try to leave untouched the dark matter sector and employ less specific scalar-tensor theories.
For instance, it has been noted in \cite{Perenon:2015sla} (see also \cite{Tsujikawa:2015mga}) that self-accelerating models within the Horndeski class with the same expansion history as $\Lambda$CDM generally supress the linear growth rate around redshift $0.5 \lesssim z \lesssim1$, despite the scalar fifth-force being attractive (see eq.~\eqref{muPhiHorn}). Looking at eq.~\eqref{deltaevol}, this can be understood by the fact that $\Omega_{\rm m}$  on the right-hand side, defined in eq.~\eqref{Omegam}, contains the time-dependent effective Planck mass $M^2$ at the denominator. The enhancement of the latter due to self-acceleration
lowers $\Omega_{\rm m}$ with respect to the standard $\Lambda$CDM case at intermediate redshifts, overcompensating $\mu_{\Phi} >1$.

As we have seen above, when the stability condition~\eqref{stabWin} is imposed the scalar force exchanged by $\pi$ in the presence of KMM is repulsive and small-scale structures are damped by a friction stronger than that provided by the Hubble expansion, see eq.~\eqref{gammaGeff}, even in the absence of self-acceleration and for a $\Lambda$CDM background expansion. In light of these facts, we consider the possibility of solving the aforementioned tension with KMM.

To illustrate this, we  compute $\sigma_8$  at redshift $z=0$ as a function of $\alpha_{\rm H,0}$ using COOP for the  cosmological parameters given at the beginning of the section. As expected from our discussion above, this yields a linear relation with $\alpha_{\rm H,0}$, i.e.
\be
\sigma_8 \simeq  (0.84 - 0.18 \, \alpha_{\rm H,0} ) \cdot \frac{A_s}{2.2 \times 10^{-9}} \;.
\label{sigma8}
\ee
In Fig.~\ref{fig:sigma8} we show this relation together with a set of large scale structure (weak lensing and cluster counts) measurements constraining $\sigma_8$.
\begin{figure}[t]
\centering
\includegraphics[width=0.9\textwidth]{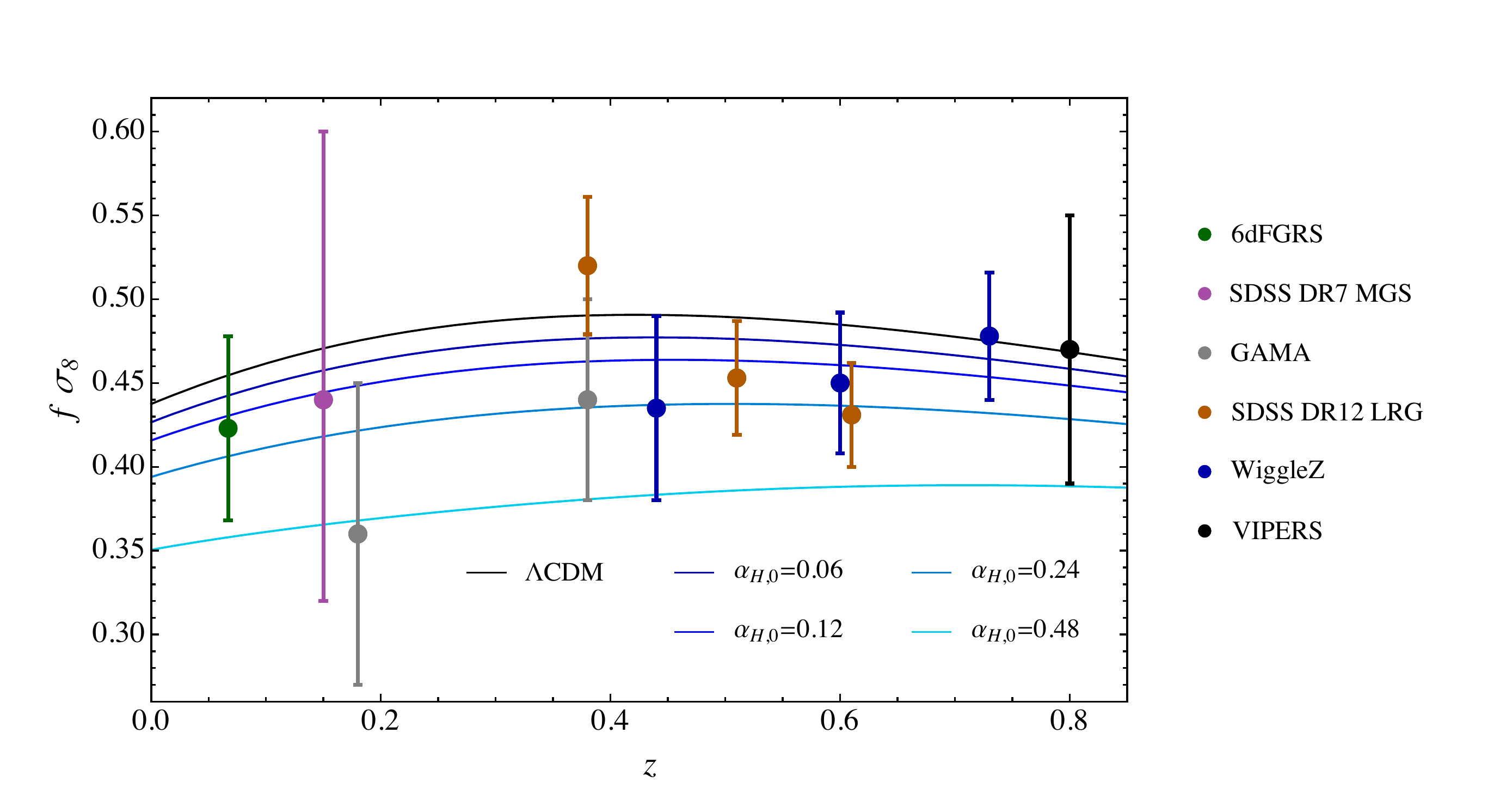}
\caption{The quantity $f \sigma_8$ as a function of redshift for different values of $\alpha_{\rm H,0}$.
The plot also shows the measurements of $f \sigma_8$ and their respective 1-$\sigma$ errors from several redshift surveys: 6dF GRS \cite{Beutler:2012px}, SDSS DR7 MGS  \cite{Howlett:2014opa}, GAMA\cite{Blake:2013nif}, SDSS DR12 LRG \cite{Alam:2016hwk}, WiggleZ \cite{Blake:2012pj} and VIPERS  \cite{delaTorre:2013rpa}. When possible, we plotted conditional constraints assuming a $\Lambda$CDM background cosmology with Planck 2015 parameters. In particular, the WiggleZ constraints were taken from Fig.~16 of~\cite{Ade:2015xua}.}
\label{fig:fsigma8}
\end{figure}
Two remarks are in order. First, it would be misleading to compute the value of $\alpha_{\rm H,0}$ that  best fits the data. Indeed, the constraints on $\sigma_8$  reported from the respective articles have been extracted from  data {\em assuming standard gravity}. Second, as explained above large values of $\alpha_{\rm H,0}$ yields superluminal scalar propagation. However, it is straightforward to choose a value of $\alpha_{\rm K,0}$ such that  the subluminality constraint \eqref{superlum} is satisfied, without affecting the redshift-survey scale evolution (as long as $c_s> 0.1$, see footnote~1).

Moreover, to illustrate the effect of KMM on the growth rate, in Fig.~\ref{fig:fsigma8} we plot the combination $f \sigma_8$ as a function of redshift for different values of $\alpha_{\rm H,0}$. Although in the presence of KMM the growth rate $f$ is scale dependent, we can confidently use its scale-independent value computed in the quasi-static regime, $f_{\rm QS}$, see eq.~\eqref{fQS}, because this approximation holds on redshift-survey scales. As discussed above, we do not try to consistently fit the value of $\alpha_{\rm H,0}$ to these observations but we note that $\alpha_{\rm H,0} \sim\text{few} \times 0.1$ would provide the hinted small-scale suppression.
A too large value of $\alpha_{\rm H,0}$ may give an unreasonably large ISW effect, see Fig.~\ref{fig:alphaH_CMB}. However, this could be compensated by a small change in another parameter, such as the dark energy equation of state. We postpone for a future publication a more consistent analysis of the CMB and large scale structure measurements  that takes into account the effects of modified gravity on the observables.

\section{Summary and conclusions}
\label{sec_last}

Using the framework of the Effective Theory of Dark Energy, in this paper we  studied the observational effects of Kinetic Matter Mixing, i.e.~a kinetic coupling between matter and the cosmological scalar field, which is present if matter is disformally coupled to the gravitational sector with a disformal coupling that depends on the first derivative of the scalar field or in theories beyond Horndeski.

In Sec.~\ref{section2}, we started by discussing the most generic quadratic action for cosmological perturbations in the presence of conformal and disformal couplings of matter to the gravitational sector, under the assumption that the disformal factor depends as well on  the first derivative of the scalar field, other than its value.
Moreover, we showed that a change of frame does not change the structure of the action but redefines the coefficients of the various operators. In particular, the coefficient of the operator that characterizes theories beyond the Horndeski class is redefined only by the dependence of the disformal coupling on the field derivative. This is explicitly shown by the frame-independent parameter $\lambda^2$, defined in eq.~\eqref{gpar}, which measures the degree of Kinetic Matter Mixing.
By diagonalizing the kinetic action, we derived the conditions that one must require for the  perturbations to be free of ghosts and of gradient instabilities (the generalization to multiple matter species is given in App.~\ref{Lagmatter}).

After this general frame-independent description, in Sec.~\ref{sec3} we assumed that matter is universally coupled and, without loss of generality, we
considered the case where it is also minimally coupled, i.e.~the Jordan frame description, where observational predictions are more easily derived.
We then discussed the short-scale regime and derived the eigenmodes of the acoustic oscillations, which are mixed states of matter and the scalar field waves. Focussing on the case where matter is made of nonrelativistic particles (such as cold dark matter or baryons)
we derived the equations  in the quasi-static approximation and discussed (see App.~\ref{app:QS}) how the quasi-static regime is reached during the cosmological evolution.
These equations allow for a clear analytical understanding of the effects of  modifications of gravity due to Kinetic Matter Mixing.
In particular, while models in the Horndeski class only modify the Poisson equation with an effective Newton constant, Kinetic Matter Mixing also induces an additional friction term. Remarkably, requiring the stability conditions implies that gravity is weakened on short scales, an effect which is hard to reproduce in models within the Horndeski class.
Finally, by comparing the quasi-static solution to the full numerical one, we showed that the quasi-static limit approximates very well the dynamics on scales shorter than the sound horizon.

In Sec.~\ref{sec:alphaH} we focussed on the cosmological effects of the beyond Horndeski operator,  obtaining the full numerical solutions using the publicly available  Einstein-Boltzmann solver of COOP~\cite{zqhuang_2016_61166}. Using these solutions, we derived the matter power spectrum at two different redshifts, and the angular spectra of the CMB lensing potential and of the CMB anisotropies.
On small scales, i.e.~for $k \gtrsim \text{few} \times 10^{-3} \textrm{Mpc}$, the solution matches the quasi-static regime and the matter power spectrum is suppressed independently of $k$. An analytical study of the large scales is complicated by the complexity of the full system of equations. However, we obtained analytical solutions on these scales by perturbing around the $\Lambda$CDM solutions for small Kinetic Matter Mixing. The agreement with the numerical solution is excellent. Moreover, its simplicity allows an immediate understanding of the behavior of the perturbations and their observables.
Similarly to the matter power spectrum, also the angular spectrum of the CMB lensing potential is suppressed.
The CMB anisotropy is affected at very low multipoles through the ISW effect, which is enhanced, and on very high multipoles because of the suppression of the lensing potential.

In App.~\ref{sec:alphaB}, we compared this case with the one of kinetic braiding, which displays qualitatively opposite effects. Also in this case we studied analytically the large-scale behavior and derived the value of the crossing scale, i.e.~the scale at which the power spectrum displays the transition between the short-scale enhancement and large-scale suppression.

As mentioned above, Kinetic Matter Mixing appears as the only modification of gravity in the context of single-field models that weakens the strength of gravity on small scales. Therefore, in Sec.~\ref{sec4.3} we entertained the possibility that the tension between the Planck data and small-scale observations can be explained by this effect. In particular, as shown in Figs.~\ref{fig:sigma8} and \ref{fig:fsigma8}, KMM predicts a lower value of $\sigma_8$ and $f\sigma_8$, which could be made compatible with those measured by weak lensing and redshift-space distortion observations. We postpone to future work a more consistent dedicated analysis that marginalizes over the other cosmological parameters.

In summary, we presented a robust theoretical understanding of the effects of Kinetic Matter Mixing across different observables and scales. These effects may be a smoking gun of modified gravity for the next observational missions and a complete forecast, taking into account the characteristics of the next missions, is an obvious next step.

\vspace{0.5cm}
\noindent
{\bf Acknowledgements:}
It is a pleasure to thank J\'er\^ome Gleyzes, Martin Kunz, David Langlois, Jean-Baptiste Melin, Federico Piazza, Ignacy Sawicki and Jeremy Tinker for useful discussions. M.M. and F.V. also thank the Theoretical Physics Departments of CERN and Universit\'e de Gen\`eve for hospitality during large part of this work. G. D'A. thanks the CCPP at New York University for hospitality during the final stages of this work.

\vspace{0.5cm}
\appendix

\section{Quadratic action and stability for multiple species}
\label{Lagmatter}

Here we study the linear stability of the gravitational and matter action, extracting the propagating degrees of freedom and their speed of propagation.
To this end, we will expand the total  action up to quadratic order in linear scalar fluctuations around a FLRW solution and solve the constraints, generalizing the treatment of \cite{Gleyzes:2015pma} by including $\alphaH$ and the dependence on $X$ of the disformal functions $D_I$. When not explicitly given, the details of the calculation can be found in this reference.

To describe the matter sector, we  extend the treatment of the main text and assume that the universe is filled by $\NS$ matter species labelled by an index $I$, with $I=1, \ldots, \NS$, each minimally coupled to a different metric. For each species $I$, we denote the corresponding metric by  $\tg^{(I)}_{\mu \nu}$ and we call this the Jordan frame metric associated with this species.
The total matter action is thus given by
\be
S_{\rm m} = \sum_I^\NS S_I \;, \qquad S_I =  \int d^4 x \sqrt{- \tg^{(I)} }\,  L_I \Big(   \tg^{(I)}_{\mu \nu}, \psi_I \Big) \; ,
\ee
with
\be
\label{disf_unit_I2}
\tg^{(I)}_{\mu \nu} = C^{(\phi)}_I(\phi) g_{\mu \nu}  + D^{(\phi)}_I(\phi, X) \partial_\mu \phi \, \partial_\nu \phi \;.
\ee
($C_I^{(\phi)}>0$ in order to preserve the Lorentzian signature of the Jordan-frame metric of the species $I$.)
As usual, one can use the arbitrariness in the choice of the gravitational metric $g_{\mu\nu}$
to choose one particular matter species, say $I_*$, to be minimally coupled to it, in which case we have $C^{(\phi)}_{I_*}=1$ and $D^{(\phi)}_{I_*}=0$.
This defines its Jordan metric as the gravitational metric.

It is convenient to introduce the parameters
\be
\label{defalphasI}
 \alphaCI \equiv \frac{\sqrt{-X}}{2 H } \frac{ d \ln C_I^{(\phi)} }{d \phi} \, , \qquad \alphaDI \equiv - \frac{X D_I^{(\phi)}}{ C_I^{(\phi)}+ X D_I^{(\phi)}}\, , \qquad \alphaXI \equiv  \frac{X^2}{ C_I^{(\phi)} } \frac{\partial D_I^{(\phi)} }{\partial X} \;  ,
 \ee
 where the right-hand side is evaluated on the background. (Requiring a Lorentzian Jordan frame metric implies $\alphaDI > -1 $ \cite{Bettoni:2013diz}.)
In unitary gauge, eq.~\eqref{disf_unit_I2}  reads
\be
\label{disf_unit_I}
\tg^{(I)}_{\mu \nu} = C_I(t) g_{\mu \nu}  + D_I(t , N) \delta_\mu^0 \delta_\nu^0 \;,
\ee
with
\be
C_I(t) =  C_I^{(\phi)} \big( \phi( t) \big)\, , \qquad D_I(t, N) =   \dot{ \phi}^2 (t) D_I^{(\phi)} \big( \phi( t) , -\dot \phi( t)^2/N^2 \big)\,.
\ee
Then, the above parameters read
\be
\label{defalphas}
 \alphaCI = \frac{\dot C_I }{2 H  C_I} \, , \qquad \alphaDI = \frac{D_I}{  C_I-D_I}\, , \qquad \alphaXI = - \frac{1}{2  C_I } \frac{\partial D_I }{\partial N} \;
 \ee

Let us start by expanding the matter action. For simplicity, we assume that each matter species can be described by a perfect fluid with vanishing vorticity. It is then easy to write an action in terms of derivatively coupled scalar fields with Lagrangians of the form \cite{ArmendarizPicon:2000dh,ArmendarizPicon:2000ah,Boubekeur:2008kn}
\be
\label{Lagmatkess}
L_I \Big(   \tg^{(I)}_{\mu \nu}, \psi_I \Big) \equiv P_I(Y_I ) \;, \qquad Y_I \equiv  \tg_{(I)}^{\mu \nu} \partial_\mu \sigma_I \partial_\nu \sigma_I \;.
\ee
Splitting   each scalar field {$\sigma_I$} into a background value and its perturbations, $\sigma_I = \bs_I (t) + \delta \sigma_I (t, \mathbf{ x}) $, the second-order expansion $S_I$ reads
 \be
 \label{actkessence}
 \begin{split}
 S^{(2)}_I  = & \int d^3 x \, dt \, \frac{ a^3 \rho_I }{\cI^2} \left\{ g_{ \delta N^2, I}   {\delta N} ^2  + \frac{1+(1+\alphaDI) w_I}{(1+\alphaDI)^2} \frac{ 1}{ 2  \dot \bs_I^2} \left[ \delta \dot \sigma_I^2 -  \cI^2 \frac{ (\partial_i \delta \sigma)^2}{a^2} \right]  \right. \\
 & \left.  - \frac{1+(1+\alphaDI) w_I}{(1+\alphaDI)^2} \frac{ 1 }{  \dot \bs_I} \left[ \delta \dot \sigma_I \left({\delta N} - \cI^2 \delta \sqrt{h}\right) + \cI^2 N^i \partial_i \delta \sigma_I \right] \right\} \;,
 \end{split}
\ee
where we have defined
\be
g_{ \delta N^2, I} \equiv \cI^2 \alphaSI+(1+\alphaXI)^2 \big[ 1 + w_I (1+\alphaDI)  \big] \;
\ee
with the combination
\be
\label{alphaSIApp}
\alphaSI \equiv \alphaDI(1+\alphaXI)^2+\alphaXI(2+\alphaXI)+ \frac{1}{2 C_I} \frac{\partial^2  D_I}{\partial N^2} \;,
\ee
and the fluid quantities
\be
\begin{split}
p_I &\equiv \frac{C_I^2}{\sqrt{1+\alphaDI}}P_I \, , \qquad \rho_I\equiv C_I^2 \sqrt{1+\alphaDI}\left(2Y_I P'_{I}-P_I\right)\, , \\
\cI^2& \equiv \frac{P'_{I}}{P'_{I} +2 Y_I  P''_{I} }\left(1+\alphaDI\right)^{-1} \;.
\end{split}
\ee
Here a prime denotes a derivative with respect to the variable $Y_I$. We have omitted in the action irrelevant terms that vanish when imposing the background equations of motion.

We can now investigate the stability of scalar perturbations.
The full second-order action
\be
\label{sec_tot_ac}
S^{(2)} = S^{(2)}_{\rm g} + S^{(2)}_{\rm m} \;,
\ee
where the gravitational part $S^{(2)}_{\rm g}$ is given in eq.~\eqref{S2},
governs the dynamics of linear scalar fluctuations.
The scalar modes can be described in unitary gauge
by defining $\psi \equiv \partial^{-2} \partial_i N^i$ and writing the spatial metric as ${h}_{ij}=a^2(t) e^{2\zeta}\delta_{ij}$ \cite{Maldacena:2002vr}.
Variation with respect to $\psi$ yields the (scalar part of) the momentum constraint,
and its  solution can be used to replace $\delta N$ in terms of $\dot \zeta$ and $\delta \sigma_I$ into the second-order action (see details in Ref.~\cite{Gleyzes:2015pma}).
Re-expressing the scalar field perturbations $\delta \sigma_I$ in terms of the gauge invariant variables
$\Q_I \equiv \delta \sigma - ({\dot \bs_I}/{H}) \zeta $,
the total second-order action reads, focusing only on the {kinetic and spatial gradient parts},
\be
\label{QuadActI}
\begin{split}
S^{(2)} = &\int d^3 x \, dt \, a^3  \, \frac{M^2}{2  }   \bigg[  g_{\dot \zeta \dot \zeta}  \dot \zeta^2 - g_{\partial \zeta \partial \zeta} \frac{(\partial_i \zeta)^2}{a^2}
+ \sum_I  \frac{\kappa_I H^2 }{\dot \bs_I^2\cI^2} \left(  \dot \Q_I^2    - \cI^2 \frac{(\partial_i \Q_I)^2}{a^2} \right) \\
& +2 \sum_I  \frac{ H}{\cI^2 \dot \bs_I}  \left(g_{\dot \Q \dot \zeta,I} {\dot \Q_I} \dot \zeta   - g_{\partial \Q \partial \zeta,I}\frac{\cI^2 }{a^2}  {\partial_i \Q_I} \partial_i \zeta \right)
\bigg] \;,
\end{split}
\ee
with
\begin{align}
g_{\dot \zeta \dot \zeta} &\equiv \frac{1}{(1+\alphaB)^2}\bigg\{\DDt+  \sum_I  \frac{\kappa_I}{\cI^2} \big[1+ \alphaB - (1+\alphaXI)(1+\alphaDI)\big]^2
\bigg\} \;, \\
g_{\partial \zeta \partial \zeta} &\equiv
  \frac{c_{s,0}^2 (\alphaK + 6 \alphaB^2 )}{(1+\alphaB)^2}+ \sum_I \frac{ \kappa_I}{1+\alphaB}  \big[1+ \alphaB- 2(1+ \alphaH)(1+\alphaDI) \big] \;, \\
g_{\dot \Q \dot \zeta,I}&\equiv  \frac{\kappa_I}{1+\alphaB}  \big[1+ \alphaB - (1+\alphaXI)(1+\alphaDI)\big] \;, \\
g_{\partial \Q \partial \zeta,I}&\equiv  \frac{\kappa_I}{1+\alphaB}    \big[ 1+\alphaB - (1+\alphaH)(1+\alphaDI)\big] \;,
\end{align}
where we defined the dimensionless coefficient
\begin{align}\label{kappa_def}
\DDt & \equiv \alphaK + 6 \alphaB^2 +3 \sum_I \alphaSI \, \Omega_I \geq0\; , \\
\kappa_I &\equiv 3 \frac{1+(1+\alphaDI) w_I}{(1+\alphaDI)^2} \Omega_I \;, \\
c_{s,0}^2 &\equiv \frac{ (1 + \alphaB)^2 }{\alphaK +6 \alphaB^2} \bigg\{ 2  (1+\alphaT) - \frac{2}{a M^2} \frac{d}{dt}\bigg[ \frac{a M^2 (1+ \alphaH) }{H(1+ \alphaB)} \bigg] \bigg\} \;. \label{cssqzero}
\end{align}
Absence of ghosts is ensured by requiring that the matrix of the kinetic coefficients is positive definite, which yields the conditions $\DDt \geq 0$ and $\kappa_I \geq 0$. The second condition reads $\rho_I+(1+\alphaDI) p_I\geq0$, which is  the usual Null Energy Condition written in a disformally related frame.

Requiring that the determinant of the kinetic matrix vanishes yields the dispersion relation
\begin{equation}
( \omega^2 -  c_s^2 k^2) \prod_{I}^{N_S} (\omega^2 - c_{s,I}^2 k^2)  =  \frac{ 3}{\alpha}  \, \omega^2 k^2  \sum_I \Big[1+(1+\alphaDI)w_{I} \Big] \Omega_{ I}\, (\alphaH-\alphaXI)^2  \prod_{J\neq I}^{N_S}
(\omega^2 - c_{s,J}^2 k^2) \, , \label{kmI}
\end{equation}
where the scalar sound speed squared $c_s^2$ is given by
\be
\label{tildecssapp}
c_s^2 \equiv  c_{s,0}^2  \frac{\alphaK +6 \alphaB^2}{\alpha} -  \frac{(1+\alphaH)^2}{\alpha} \sum_I  \kappa_I (1+\alphaDI)^2 \;.
\end{equation}
For a single matter fluid this yields eq.~\eqref{km}.
In the absence of a  disformal coupling, $\alphaDI= \alphaXI =0$, we recover the results of \cite{Gleyzes:2014dya,Gleyzes:2014qga}. If the disformal coupling does not depend on $X$, $\alphaXI =0$, and we restrict to Horndeski theories, $\alphaH=0$, we recover the results of \cite{Gleyzes:2015pma}.

\section{Quadratic action in Newtonian gauge}
\label{Fullaction}

The second-order action \eqref{S2} can be written in Newtonian gauge, eq.~\eqref{Newtoniangauge}, after a time diffeomorphism $t \to t +\pi (t, \vec x)$. This reads
\be
\begin{split}
\label{fullacgrav}
S_{\rm grav}^{(2)}=& \!\int \!d^4x a^3M^2 \bigg\{ \frac12 H^2 \alphaK \pid^2+\bigg[\dot H  + \frac1{2M^2} \big(\rho_{\rm m} + p_{\rm m} +2(M^2 H (\alphaB -\alphaH) )^{\hbox{$\cdot$}}  \big)\\ & +  H^2 (\alphaB - \alphaM + \alphaT - \alphaH)      \bigg]  \frac{(\gpi)^2}{a^2}
-3 \dot \Psi^2 + (1+\alphaT) \frac{(\nabla\Psi)^2}{a^2}  -2(1+\alphaH) \frac{\nabla\Phi \nabla\Psi}{a^2}
 \\
& +2 \alphaH \frac{\nabla\pid \nabla \Psi}{a^2}  +2 H ( \alphaB - \alphaH )\nabla\Phi\gpi -2 H (\alphaM -\alphaT) \frac{ \nabla\Psi \gpi}{a^2}
+6 H  \alphaB\pid \dot \Psi \\
&+H^2 ( 6 \alphaB-\alphaK ) \Phi \pid   - 6 H(1+ \alphaB) \dot \Psi \Phi -6\left(\frac{\rho_{\rm m}+p_{\rm m}}{2 M^2}+\dot{H}\right)\dot{\Psi} \pi  \\
& + \bigg[H^2 \left( \frac12 \alphaK - 3 (1 +2 \alphaB)  \right)+\frac{\rho_{\rm m}}{2M^2} \bigg]\Phi^2 -  \frac{ 9 p_{\rm m} }{2 M^2}  \Psi^2-\frac{3 \rho_{\rm m}}{M^2} \Phi \Psi \\
& -3 H\left(\frac{\rho_{\rm m}+p_{\rm m}}{2 M^2}+\dot{H}(1+\alphaB)\right) \Phi \pi \\
& -3 \bigg[\dot{H}\left(\frac{\rho_{\rm m}+p_{\rm m}}{2 M^2}+\dot{H}\right)+H (\alphaB \dot{H})^{\hbox{$\cdot$}}+H^2 \dot{H} \alphaB (3+\alphaM)+\alphaB {\dot{H}}^2\bigg] \pi^2 \bigg\} \;  .
\end{split}
\ee
In Newtonian gauge, the matter action \eqref{Smatter}  expanded at second-order  reads
\be
\label{fullacmat}
\begin{split}
S_{\rm m}^{(2)} =
&
\int d^4x \; a^3
\bigg\{ \frac{9}{2}
p_{\rm m} \Psi^2
+\frac{1}{2} \bigg[ \frac{p_{\rm m}
+(1-c_{\rm m}^2) \rho_{\rm m} }{c_{\rm m}^2}
\bigg]\Phi^2+3 \rho_{\rm m} \Phi \Psi \\
&+\frac{ \rho_{\rm m}+p_{\rm m}}{2 c_{\rm m}^2 \dot{\sigma}_0^2}\bigg[ \dot{\delta \sigma}^2- c_{\rm m}^2\frac{{(\nabla \delta \sigma)}^2}{a^2}\bigg] -\frac{ \rho_{\rm m}+p_{\rm m}}{c_{\rm m}^2 \dot{\sigma}_0}\,(\Phi+3 c_{\rm m}^2\Psi) \, \dot{\delta \sigma} \bigg\} \; .
\end{split}
\ee

\section{From the oscillating to the quasi-static regime}
\label{app:QS}

We can discuss the transition from the oscillating to the quasi-static regime starting from the linearized equations of motion for $\Psi$ in the presence of matter, once the $\pi$ field has been integrated out \cite{Bellini:2014fua,Gleyzes:2014rba,Lombriser:2015cla}.
On short scales, in Fourier space the variation of the action with respect to $\Psi$ and the relation $\Phi_E = \Psi_E$  read, respectively,
\begin{align}
\label{oscipsi}
 \ddot{\Psi} + (3+b_1) H \dot{\Psi} +  c_s^2 \frac{k^2}{a^2}  \Psi
&= - \frac{3}{2} \Om_{\rm m} H^2 \[  c_s^2 \mu_{\Psi} \delta_{\rm m} - \frac{2 \alphaH (\alphaB - \alphaH)}{\al}  \frac{k^2}{a^2 H^2} H v_{\rm m} \] \; ,\\
\frac{\alphaH}{{\alphaB - \alphaH}}   \frac{\dot{\Psi}}{H} +  b_2  \Psi -  \Phi
&= - \frac32 \Om_{\rm m} \frac{a^2 H^2}{k^2} b_3
\delta_{\rm m} + \frac{3}{2} \frac{\alphaH}{\alphaB - \alphaH} \Om_{\rm m} H v_{\rm m} \;,\label{psiphi}
\end{align}
where the specific form of the coefficients $b_1$, $b_2$ and $b_3$ are explicitly given by
\begin{align}
b_1 & \equiv 3 + \alphaM + \frac{\alphaB^2}{H\alpha } \left( \frac{\alphaK}{\alphaB^2} \right)^{\hbox{$\cdot$}} + \alphaH \left[  \frac2{\alphaH -\alphaB} \left( \frac{\dot {\alpha}_{\rm B}}{H \alphaB} - \frac{\dot {\alpha}_{\rm H}}{H \alphaH}\right) + \frac34 \Omega_{\rm m} \right] \;, \\
b_2 & \equiv \frac{1}{(\alphaB -\alphaH)^2} \left[ \alphaB \xi +  \alphaH(1+\alphaH) \bigg( \frac{\dot H}{H^2} + \frac32 \Omega_{\rm m} \bigg)  \right]\;, \\
b_3 & \equiv  \frac{ 1}{
(\alphaB -\alphaH)^2} \left[ {\alphaB}  (\alphaT - \alphaM)    +  \alphaH  \bigg(\frac{\dot H}{H^2} + \frac{3}{2} \Omega_{\rm m} \bigg) \right]\; .
\end{align}
In the standard quintessence case, for $\alphaH = \alphaT=\alphaM=\alphaB=0$, we have  $b_1 = 0$, $b_2=1$ and $b_3=0$.
Matter is described by the usual continuity and Euler equations, eqs.~\eqref{conti} and \eqref{Euler}.

One cannot find an analytical solution to eqs.~\eqref{oscipsi} and \eqref{psiphi} but we can assume that  the full solution can be separated
into an oscillating part, with characteristic frequency $\om \sim c_+ k/a =  c_s \sqrt{1 + \lambda^2}\, k/a$, and a ``quasi-static'' part, slowly evolving at a rate given by $\sim H$ \cite{Sawicki:2015zya}. For instance, for $\Psi$ one can write
\be
\Psi = \Psi_{\rm osc} +  \Psi_{\rm QS} \;.
\ee
We also assume that $\Psi_{\rm osc}$ has a slowly decaying envelop due to the expansion of the universe, such that
\be
\dot \Psi = \dot f \Psi_{\rm osc} + i \omega \Psi_{\rm osc} + \dot \Psi_{\rm QS} \;, \qquad \dot f \sim {\cal O}(H) \;, \qquad  ( \ln \Psi_{\rm QS} )^{\hbox{$\cdot$}}\sim {\cal O}(H)  \;.
\ee
To count the importance of each term in the above equations, we consider the limit $\omega \gg H$ and we define the  following two expansion parameters,
\be
\epsilon_k \equiv \frac{aH}{k} \ll 1 \;, \qquad \epsilon_\omega \equiv \frac{H}{\omega} = \frac{a H}{c_+ k} \ll1 \;.
\ee
Starting by defining $\Psi \sim \OO(1)$, and using eq.~\eqref{oscipsi} and the continuity and Euler equations, one can find
\be
\begin{split}
&H v_{\rm QS} \sim \OO (1)\;, \qquad \Phi_{\rm QS} \sim \OO ( 1) \;,  \qquad \delta_{\rm QS} \sim \OO  (\epsilon_k^{-2}  )\;, \\
&H v_{\rm osc} \sim \OO (1)\;, \qquad \Phi_{\rm osc} \sim \OO ( \epsilon_\omega^{-1}) \;,  \qquad \delta_{\rm osc} \sim \OO  (\epsilon_k^{-2} \epsilon_\omega )
\;.
\end{split}
\ee

At this point, we can expand the above equations in these expansion parameters. At the lowest order in $\epsilon_k$, the quasi-static solutions satisfy the relations discussed in
Sec.~\ref{sec:QS},
not surprisingly. For the oscillating piece,
eqs.~\eqref{oscipsi} and \eqref{psiphi} become, retaining only the lowest order in $\epsilon_\omega$ and $\epsilon_k$,
\begin{align}
- \om^2 \Psiosc +  c_s^2 \frac{k^2}{a^2} \Psiosc
&\simeq 3 \Om_{\rm m} \frac{\alphaH (\alphaB - \alphaH) }{\al} \frac{k^2}{a^2} H \vosc \; , \\
\frac{\alphaH}{{\alphaB - \alphaH}} i \frac{\om}{H} \Psiosc
- \Phiosc & \simeq 0 \;, \label{constr}
\end{align}
where we have used  $\dot f /H \simeq -(3 + b_1)  + \OO ( \epsilon_{\omega})$. For $ b_1 > - 3$, this implies that the oscillating part decays in time. For instance, for a constant $b_1$ the oscillating solution decays as $a^{-(3+b_1)}$.
Combining the Euler equation $i \om \vosc = - \Phiosc$ with eq.~\eqref{constr} we get a simple relation between the velocity and the curvature, i.e.~$H \vosc \simeq - {\alphaH}/({\alphaB - \alphaH}) \Psiosc$.
Replacing this expression in the first equation, we find the expected dispersion relation for the oscillating normal mode,
\be
\om^2 = c_s^2 \frac{k^2}{a^2} + 3 \Om_{\rm m} \frac{\alphaH^2}{\al} \frac{k^2}{a^2} = c_+^2 \frac{k^2}{a^2} \, .
\ee

\section{Observational signatures of Kinetic Braiding}
\label{sec:alphaB}

It is interesting to compare the results of
Sec.~\ref{sec:alphaH}
with the
case
of kinetic braiding. Indeed, this modification of gravity is expected to lead to similar effects as KMM on the power spectrum and CMB anisotropies.
We assume the same background expansion history as in eq.~\eqref{Hparametr} and  set, this time,
\be
\alphaH=\alphaM=\alphaT=0 \;.
\ee
Moreover, we parametrize the time dependence of
$\alphaK$ and $\alphaB$  as
\be
\alphaK =  \alpha_{\rm K,0} \frac{\Omega_{\rm DE} (t) }{\Omega_{\rm DE,0}}\;, \qquad
\alphaB =  \alpha_{\rm B,0} \frac{\Omega_{\rm DE} (t) }{\Omega_{\rm DE,0}}\;.
\ee
Recently, in Ref.~\cite{Zumalacarregui:2016pph} an analogous parametrization has been
used to discuss the effect of $\alphaB$---as well as of other parameters---on the power spectrum and the CMB anisotropies (see also \cite{Gleyzes:2015pma,Renk:2016olm}).
We agree with the results of Fig.~2 of this reference, for the corresponding values of $\alpha_{\rm B,0}$ and $\alpha_{\rm K,0}$.\footnote{We can compare with Ref.~\cite{Zumalacarregui:2016pph} by the following correspondence between our parameters $\alpha_{\rm B,0}$ and $\alpha_{\rm K,0}$ and their parameters $\hat \alpha_{\rm B}$ and $\hat \alpha_{\rm K}$: $\alpha_{\rm B,0} = -  \hat \alpha_{\rm B} \Omega_{\rm DE,0}/2$ and $\alpha_{\rm K,0} =  \hat \alpha_{\rm K} \Omega_{\rm DE,0}$.}

From the above assumptions it follows that
the speed of scalar fluctuations is
\be
c_s^2 = - \frac{\alphaB  \big[ 2 (1+ \alphaB)  + 3 \Omega_{\rm m}  \big]}{\alphaK + 6 \alphaB^2} \;.
\ee
Absence of ghosts and gradient instabilities therefore imply respectively that
$\alphaK  +6 \alphaB^2 \ge 0$
and $- 1 -(3 /2)\Omega_{\rm m}  \le \alphaB \le 0$.
As in Sec.~\ref{sec:alphaH}, we set  $\alpha_{\rm K,0} =1$; then we study the effect of $\alphaB$  for four negative values of $\alpha_{\rm B,0}$ to avoid instabilities: $\alpha_{\rm B,0} = - 0.06$, $- 0.12$, $- 0.24$ and $- 0.48$.

\subsection{Matter power spectrum}

\begin{figure}
\centerline{\includegraphics[width=0.505\textwidth]{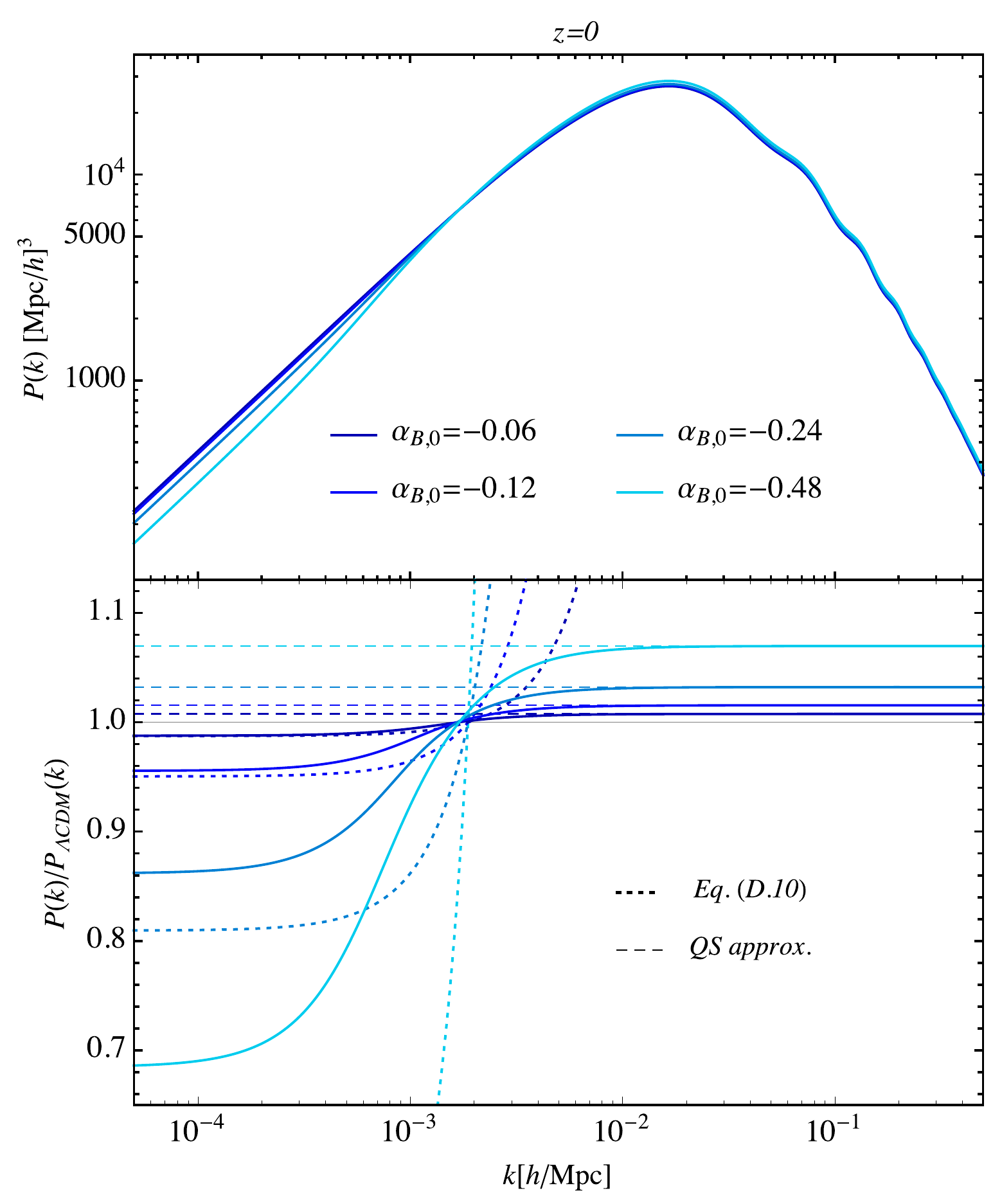}
\includegraphics[width=0.5\textwidth]{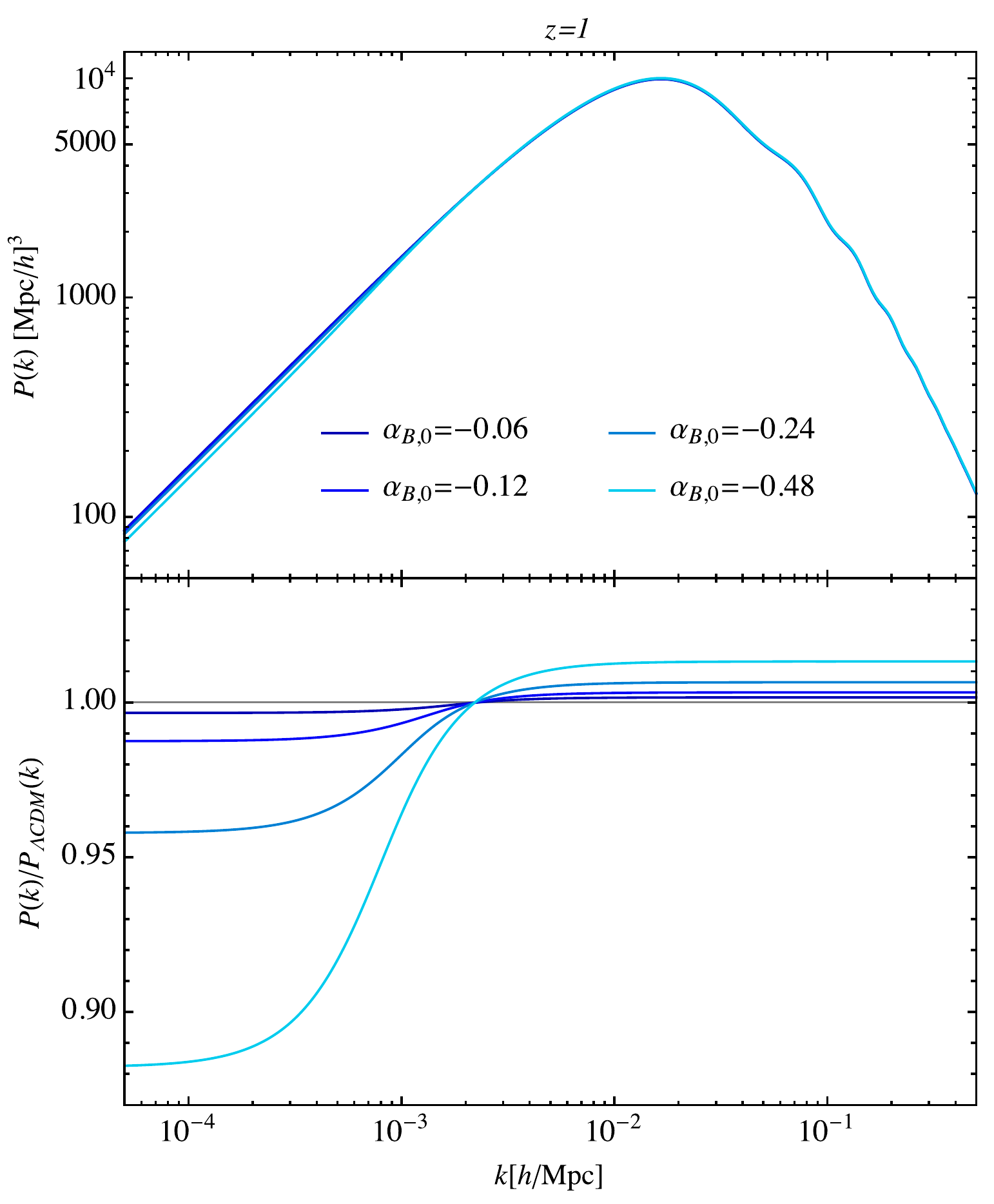}}
\caption{Matter power spectrum for four different values of $\alphaB$ today, i.e.~$\alpha_{\rm B,0} = -0.06$, $-0.12$, $-0.24$ and $-0.48$, at redshift $z=0$ (left panel) and $z=1$ (right panel).
For comparison, the dashed and dotted lines in the left lower panel respectively show the quasi-static approximation and the perturbative solution of eq.~\eqref{DeltaB}.}
\label{fig:alphaB_PS}
\end{figure}

We plot the matter power spectrum in Fig.~\ref{fig:alphaB_PS}. The effect of $\alphaB$ is to enhance the power on short scales, due to strengthening of gravity. Indeed, the modification of the Poisson equation \eqref{Geff} reads
\be
\mu_\Phi = 1-\frac{\alphaB}{1+\alphaB +3 \Omega_{\rm m}/2 } \;,
\ee
and one can use this relation in eq.~\eqref{deltaevol} (with $\gamma=0$) to predict the corresponding enhancement. On large scales we observe the opposite effect, i.e.~a suppression of power, and a crossover scale between these two regimes independent of $\alpha_{\rm B,0}$.

To study the large scale regime we proceed analogously to what done in
Sec.~\ref{sec:mps}
and solve the Einstein equations perturbatively in $\alphaB$. In this case, the relevant equations are
\begin{align}
 \dot \Phi + H \Phi - \dot H v_{\rm m} &= \alphaB H (\dot \pi - \Phi)\;, \label{E2B}\\
 - \frac{k^2}{a^2} \Phi + \dot H \Delta_{\rm m} - \frac{\alphaK}2  H^2 (\dot \pi - \Phi)& = \alphaB H \bigg[ 3 \dot \Phi + 3 H \Phi + 3 \dot H \pi - \frac{k^2}{a^2} H \pi \bigg] \label{E1B} \;,\\
\ddot \Phi + 4 H \dot \Phi  + (2 \dot H + 3 H^2) \Phi & = \alphaB \big[ \ddot \pi - \dot \Phi + (3H^2 - \dot H) ( \dot \pi - \Phi) \big] \;, \label{E4B} \\
 \frac{ 1}{2}(\alphaK + 6 \alphaB^2) H^2 a^{-3} \frac{d}{dt} \left[ a^3 (\dot \pi - \Phi) \right]   &=  \alphaB  \bigg\{  \frac{k^2}{a^2} \big[ H\Phi + (H^2- \dot H) \pi  \big]  \nonumber \\
 & \quad +  3 \dot H \big[ 2 \dot \Phi + 2 H \Phi +  \dot H (\pi - v_{\rm m}  ) \big]    \bigg\} \;. \label{EpiB}
\end{align}
Equation \eqref{E2B} is the ``$0i$'' scalar component  of the Einstein equations, eq.~\eqref{E1B} follows from combining the ``$00$'' component with eq.~\eqref{E2B},  eq.~\eqref{E4B} is the trace of  the  ``$ij$'' components and we have used $\Psi = \Phi$, which follows from the traceless part of  the  ``$ij$'' components. Finally, eq.~\eqref{EpiB} is the evolution equation for $\pi$.

For $\alphaB=0$, these equations have the solution given in eqs.~\eqref{solalphaHzero} and $\pi = -\epsilon$. As in the case of $\alphaH$, $\alphaB$ does not affect the metric and matter perturbations at first order: eq.~\eqref{solalphaHzero} remains a solution with
\be
\pi = - \epsilon - 2 \frac{\alphaB}{\alphaK} k^2 \int \frac{\epsilon dt}{a^2 H} + {\cal O}(\alphaB^2) \;.
\ee

In order to see the effects of braiding we need to go at second order in $\alphaB$ \cite{Creminelli:2008wc}. The matter density contrast $\Delta_{\rm m}$ can be computed from eq.~\eqref{E1B}, similarly to what discussed in Sec.~\ref{sec:mps}.  We can solve for $\Phi$ at second order from eq.~\eqref{E4B}, where we use the first-order solution on the right-hand side. To derive $\dot \pi -\Phi$, we can solve eq.~\eqref{EpiB} after replacing $v_{\rm m} $ using \eqref{E1B}.
In conclusion, the density constrast reads
\be
 \Delta_{\rm m}   = \Delta_{{\rm m}, \Lambda\text{CDM} } \left[1 - 2  \frac{\alphaB^2}{\alphaK}  \left( F_1 -  \frac{k^2}{a^2 H^2} F_2 \right) \left(1 - \frac{H}{a} \int a dt  \right)^{-1} + {\cal O}(\alphaB^4)\right]  \;,
 \label{DeltaB}
\ee
where
\be
\begin{split}
F_1&\equiv \frac{3H^2}{a} \left( \frac{1}{H} \int a dt - \int \frac{a dt}{H}  \right) \;, \\
F_2& \equiv a H^4 \left[  H \int  \frac{1}{a^2 H^2} \left( \frac{2}{H} \int a dt - \int \frac{a dt}{H}  \right)  -  \int \frac{dt}{a^2 H^2} \int a dt\right] \;.
\end{split}
\ee
Equation \eqref{DeltaB} explains the large scale suppression in the power spectrum and why the crossover scale, which can be derive from the above equation as
\be
k_{\rm c} = a H \sqrt{{F_1}/{F_2}} \;,
\ee
is independent of $\alphaB$.
However, in  Fig.~\eqref{fig:alphaB_PS}  we observe a large discrepancy between  eq.~\eqref{DeltaB} and the output of COOP. It can be checked that the difference grows as $\alphaB^4$ and it is thus  due to the neglected corrections to eq.~\eqref{DeltaB}.

\subsection{Cosmic microwave background}

\begin{figure}
\centerline{\includegraphics[width=0.496\textwidth]{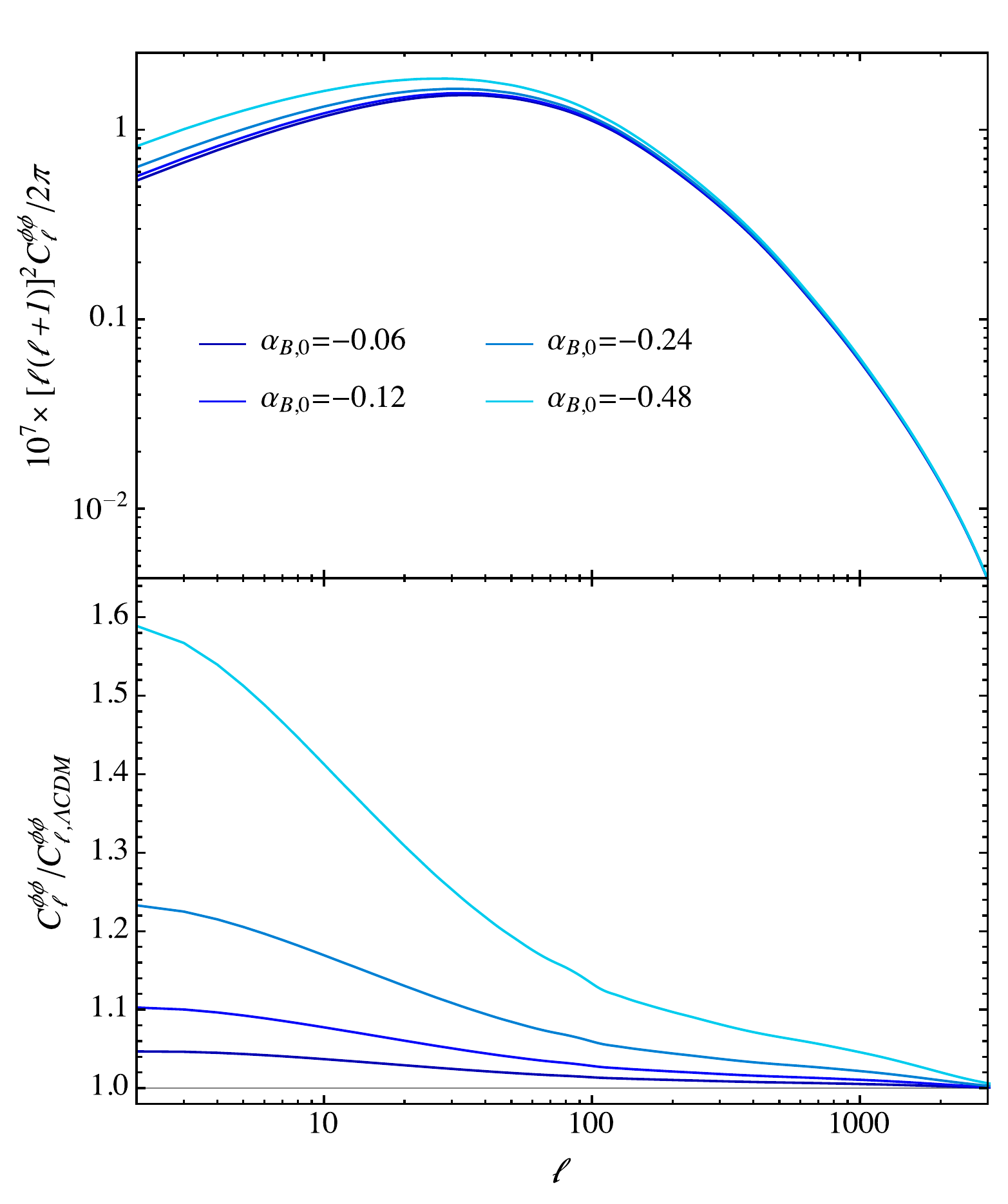}
\includegraphics[width=0.5\textwidth]{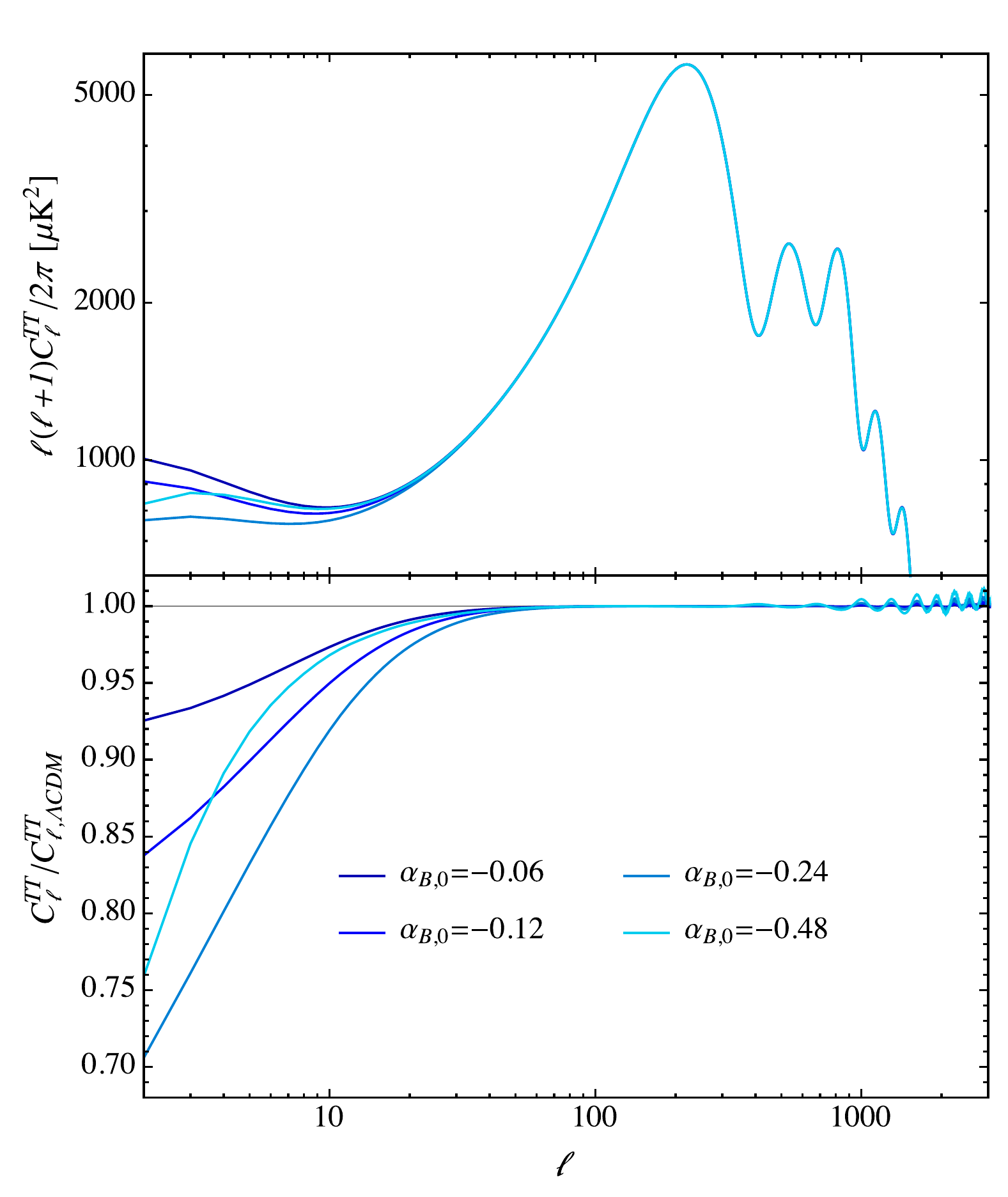}}
\caption{Effect of braiding ($\alphaB$) on the CMB lensing potential  (left panel) and on the CMB anisotropies (right panel) angular power spectra. The lower plots display the ratio of these angular spectra with the respective spectra for $\alphaB=0$.}
\label{fig:alphaB_CMB}
\end{figure}

In Fig.~\ref{fig:alphaB_CMB} we plot the angular power spectrum of the lensing potential (left panel) and of the CMB anisotropies (right panel). A negative braiding parameter $\alphaB$ induces an enhancement in the lensing potential. Similarly to what done in the previous section, we can understand this effect as a modification of the Weyl potential, expressed in terms of the parameter $\mu_{\rm WL}$ in eq.~\eqref{lensingmu}. Setting $\alphaM=\alphaT=\alphaH=0$, this reads (see also \cite{Gleyzes:2015rua} for an analysis using the quasi-static approximation)
\be
\mu_{\rm WL} -2 =  - \frac{2 \alphaB}{1+ \alphaB + 3 \Omega_\tm /2} \;.
\ee
This relation shows that  for negative values of $\alphaB$, the Weyl potential is enhanced for all redshifts. Comparing with the effect of $\alphaH$ shown in Fig.~\ref{fig:alphaH_CMB}, we notice that here the effect is larger at smaller $l$; this is due to the fact that, contrarily to the $\alphaH$ case, here $\mu_{\rm WL} -2$ does not change sign at low redshift, and contributes also to low multipoles.

Let us turn now to the CMB angular power spectrum, right panel of Fig.~\ref{fig:alphaB_CMB}. Increasing $-\alphaB$ enhances the lensing potential, thus increasing the smearing effect on the CMB acoustic peaks, as shown on the right  lower panel. The suppression of  the ISW effect can be understood again by looking at eq.~\eqref{ISW}. Now
\be
\frac{d \ln \mu_{\rm WL}}{d \ln a} = - \frac{15 \alphaB \Omega_\tm}{(2+3 \Omega_\tm) (1+\alphaB + 3 \Omega_\tm/2)} \;,
\ee
which for negative values of $\alphaB$ is positive, i.e.~has opposite sign as the standard $\Lambda$CDM contribution coming from the first two terms in the right-hand side of eq.~\eqref{ISW}. For small values of $-\alpha_{\rm B,0}$, $ d \ln \mu_{\rm WL}/ d \ln a$ is smaller than $1- d \ln \delta_\tm/d \ln a $: the time derivative of the Weyl potential remains negative and the net ISW effect is suppressed by
kinetic braiding.
For large values of $-\alpha_{\rm B,0}$, i.e.~$- \alpha_{\rm B,0} \gtrsim 0.3$, the right-hand side of eq.~\eqref{ISW} changes sign and increasing $\alphaB$ enhances the ISW effect.

 \bibliographystyle{utphys}
\bibliography{EFT_DE_biblio3}

\end{document}